\documentclass[fleqn,usenatbib]{mnras}
\usepackage{graphicx}	
\usepackage{amsmath}

\DeclareRobustCommand{\VAN}[3]{#2}
\let\VANthebibliography\thebibliography
\def\thebibliography{\DeclareRobustCommand{\VAN}[3]{##3}\VANthebibliography}

\usepackage[varg]{txfonts}
\usepackage{subcaption}
\usepackage{natbib}
\usepackage{multirow}
\usepackage{comment}
\usepackage{xcolor}
\bibpunct{(}{)}{;}{a}{}{,}

\title[Hydrogen emission from meteors and meteorites]{Hydrogen emission from meteors and meteorites: mapping traces of H\textsubscript{2}O molecules and organic compounds in small Solar system bodies}

\author[P. Matlovič et al.]
{Pavol Matlovič$^{1}$\thanks{E-mail: matlovic@fmph.uniba.sk},
Adriana Pisarčíková$^{1}$,
Juraj Tóth$^{1}$,
Pavel Mach$^{1}$,
Peter \v{C}ermák$^{1}$,
Stefan Loehle$^{2}$,
\newauthor{Leonard Kornoš$^{1}$,
Ludovic Ferri\`ere$^{3}$,
Jiří Šilha$^{1}$,
David Leiser$^{2}$,
Ranjith Ravichandran$^{2}$}
\\
$^{1}$Faculty of Mathematics, Physics and Informatics, Comenius University in Bratislava, Mlynska dolina, 84248 Bratislava, Slovakia\\
$^{2}$High Enthalpy Flow Diagnostics Group, Institute of Space Systems, University of Stuttgart, 70569 Pfaffenwaldring, Stuttgart, Germany\\
$^{3}$Natural History Museum Vienna, Burgring 7, 1010 Vienna, Austria\\
}

\date{Accepted for publication in Monthly Notices of the Royal Astronomical Society, March 31st, 2022}
\pubyear{2022}

\begin{document}
\label{firstpage}
\pagerange{\pageref{firstpage}--\pageref{lastpage}}
\maketitle

\begin{abstract}
The hydrogen emission from meteors is assumed to originate mainly from the meteoroid composition, making it a potential tracer of H\textsubscript{2}O molecules and organic compounds. H$\alpha$ line was previously detected in individual fireballs, but its variation in a larger meteor dataset and dependency on the dynamical origin and physical properties have not yet been studied. Here we investigate the relative intensity of H$\alpha$ within 304 meteor spectra observed by the AMOS network. We demonstrate that H$\alpha$ emission is favored in faster meteors ($v_i >>$ 30 km s\textsuperscript{-1}) which form the high-temperature spectral component. H$\alpha$ was found to be a characteristic spectral feature of cometary meteoroids with $\sim$ 92\% of all meteoroids with detected H$\alpha$ originating from Halley-type and long-period orbits. Our results suggest that hydrogen is being depleted from meteoroids with lower perihelion distances (q $<$ 0.4 au). No asteroidal meteoroids with detected H emission were found. However, using spectral data from simulated ablation of different meteorite types, we show that H emission from asteroidal materials can occur, and apparently correlates with their water and organic matter content. Strongest H emission was detected from carbonaceous chondrites (CM and CV) and achondrites (ureilite and aubrite), while it was lacking in most ordinary chondrites. The detection of H$\alpha$ in asteroidal meteoroids could be used to identify meteoroids of carbonaceous or achondritic composition. Overall, our results suggest that H$\alpha$ emission correlates with the emission of other volatiles (Na and CN) and presents a suitable tracer of water and organic matter in meteoroids.
\end{abstract}

\begin{keywords} 
meteorites, meteors, meteoroids -- astrobiology -- surveys -- techniques: spectroscopic
\end{keywords}

\section{Introduction} \label{sec:intro}

Dynamical models have shown that water-rich asteroids, comets and their fragments - meteoroids, can deliver water to objects throughout the solar system. Meteorite analyses of angrites and eucrites \citep{2014Sci...346..623S, 2017GeCoA.212..156S} suggest the accretion of water in the inner solar system within the first few million years of its existence. Based on the present isotopic evidence, impactors with composition similar to carbonaceous chondrites were the main deliverers of water and volatiles in the inner solar system \citep{2013Sci...340.1317S, 2014Sci...346..623S, doi:10.1126/sciadv.abg9707}. Recent measurements of hydrogen contents and deuterium/hydrogen ratios also point out that most of Earth's water could have formed from hydrogen delivered by enstatite chondrite meteorites \citep{doi:10.1126/science.aba1948}.

Meteor observations allow us to continuously study the properties of small interplanetary bodies that impact the Earth’s atmosphere today. Many of these meteoroids are already linked to a meteoroid stream, often originating from a known parent comet or asteroid. By studying the spectral features of the incoming meteors and computing their original heliocentric orbits, we are then also able to infer compositional information about the parent comets and asteroids, including their past volatile content.

It was proposed that the increased content of water and organics in meteoroids can be traced by observing the hydrogen emission in meteors. \citet{2004AsBio...4..123J} have shown that the H emission in meteoric plasma must dominantly originate in the meteoroid composition and not from the Earth's atmosphere, making it a suitable diagnostic tool for investigating the hydrogen content in meteoroids. The source of this hydrogen is assumed to be from ablating refractory organic matter and/or from H\textsubscript{2}O molecules trapped and bound in minerals (i.e. OH- or H\textsubscript{2}O in silicate minerals, and C-bonded H in organic compounds). The loss of a significant fraction of hydrogen from meteoritic organic matter occurs during ablation, possibly in a process known as carbonization, in which functional groups are lost.

Two hydrogen lines have been recognized in meteor spectra. The H$\beta$ line near 486.1 nm is typically weaker and can be lost in the spectrum noise for some fainter meteors. A focus is therefore placed on the stronger H$\alpha$ line, which can be recognized in a wider range of meteor magnitudes. H$\alpha$ has high excitation energy of 12.09 eV, so it is expected to be weaker when the line is thermally excited at around 4400 K -- the temperature characteristic for the low-temperature component of the meteor plasma \citep{1994P&SS...42..145B}. It is assumed that a high-temperature component (T $\approx$ 10 000 K) can form in plasma of fast meteors, in which a stronger H$\alpha$ line can be observed, along with other specific features such as the first-degree ionized Ca II and Si II.

The detection of the H$\alpha$ line in meteor spectra was previously reported in several individual fireball observations \citep{1971CaJPh..49.1738C, 2000EM&P...82..399B, 2004AsBio...4..123J, 2019A&A...629A..71M, 2019A&A...629A.137S}, but an analysis of its variations within a larger dataset of meteors has not yet been performed. H abundances in Leonid and Perseid meteoroids were analyzed by \citet{2004EM&P...95..245B}, who found enhancement of H abundance in comparison with CI chondrites. In this work, we study the occurrence and relative intensity of H$\alpha$ in a large sample of 200 mm- to dm- sized meteoroids with determined orbital properties. By analyzing the H$\alpha$ emission in a large dataset of mid-sized meteoroids, we aim to identify the characteristic properties of meteoroids with detected hydrogen content, reveal the variations of H emission in meteoroids from different dynamical sources, and study the implications of the increased H content on the meteoroid material strength. Additionally, we utilize the performed experiments of laboratory meteorite ablation in plasma wind tunnel to study the potential H emission from asteroidal samples of different composition.

First, in Section \ref{sec:methods}, we describe the observations and methods applied in this analysis. Our results (Section \ref{sec:results}) are divided into an analysis of the dynamical and physical properties of meteoroids with detected H emission (Section \ref{sec:survey}), and an analysis of the variation of H emission in spectra of ablated meteorite samples of different types (Section \ref{sec:laboratory}). Finally, the discussion and conclusions of the obtained results are provided in Sections \ref{sec:discussion} and \ref{sec:conclusions}, respectively.

\section{Observations and methods} \label{sec:methods}

The meteor and meteor spectra observations used in this work were captured by the global All-sky Meteor Orbit System (AMOS) network. The AMOS network, operated by the Comenius University in Bratislava, currently consists of 14 standard AMOS systems providing trajectory and orbital data \citep{2015P&SS..118..102T,Toth2019AMOST} and 8 spectral systems AMOS-Spec providing spectral data \citep{2016P&SS..123...25R, 2019A&A...629A..71M,2020A&A...636A.122M}. AMOS systems are located at five stations in Slovakia, two stations in the Canary Islands, two stations in Chile, two stations in Hawaii and three stations in Australia.

The analysis presented here is based on a survey of 304 meteor spectra observations (a sample expanded from the survey presented in \citet{2019A&A...629A..71M}), with focus placed on 236 of these events, for which the orbital and material strength properties were determined. The majority (200) of these selected events were captured by the lower resolution AMOS-Spec system in Slovakia during the 2013-2019 period. In addition, 36 observations with confirmed H$\alpha$ emission were selected for this study from the higher resolution system AMOS-Spec-HR in Hawaii (Maunakea observatory) and Canary Islands (Roque de los Muchachos and Teide observatories), captured during the 2016-2021 period.

\subsection{Instrumentation}

The AMOS system consists of four major components: a fish-eye lens, an image intensifier, a projection lens, and a digital video camera. The resulting field of view (FOV) of AMOS is 180$^{\circ}$x140$^{\circ}$ with an image resolution 1600x1200 (20 fps) for stations in the Canary Islands, Hawaii, Chile and Australia, and 1280x960 (15 fps) pixels for stations in Slovakia. This translates to a resolution of 6.8 and 8.4 arcmin pixel$^{-1}$, respectively. Limiting magnitude for stars is around +5 mag for a single frame. The detection efficiency is lower for moving objects due to the trailing loss, approximately +4 mag at typical meteor speeds.

\begin{figure}
\centerline{\includegraphics[width=.90\columnwidth,angle=0]{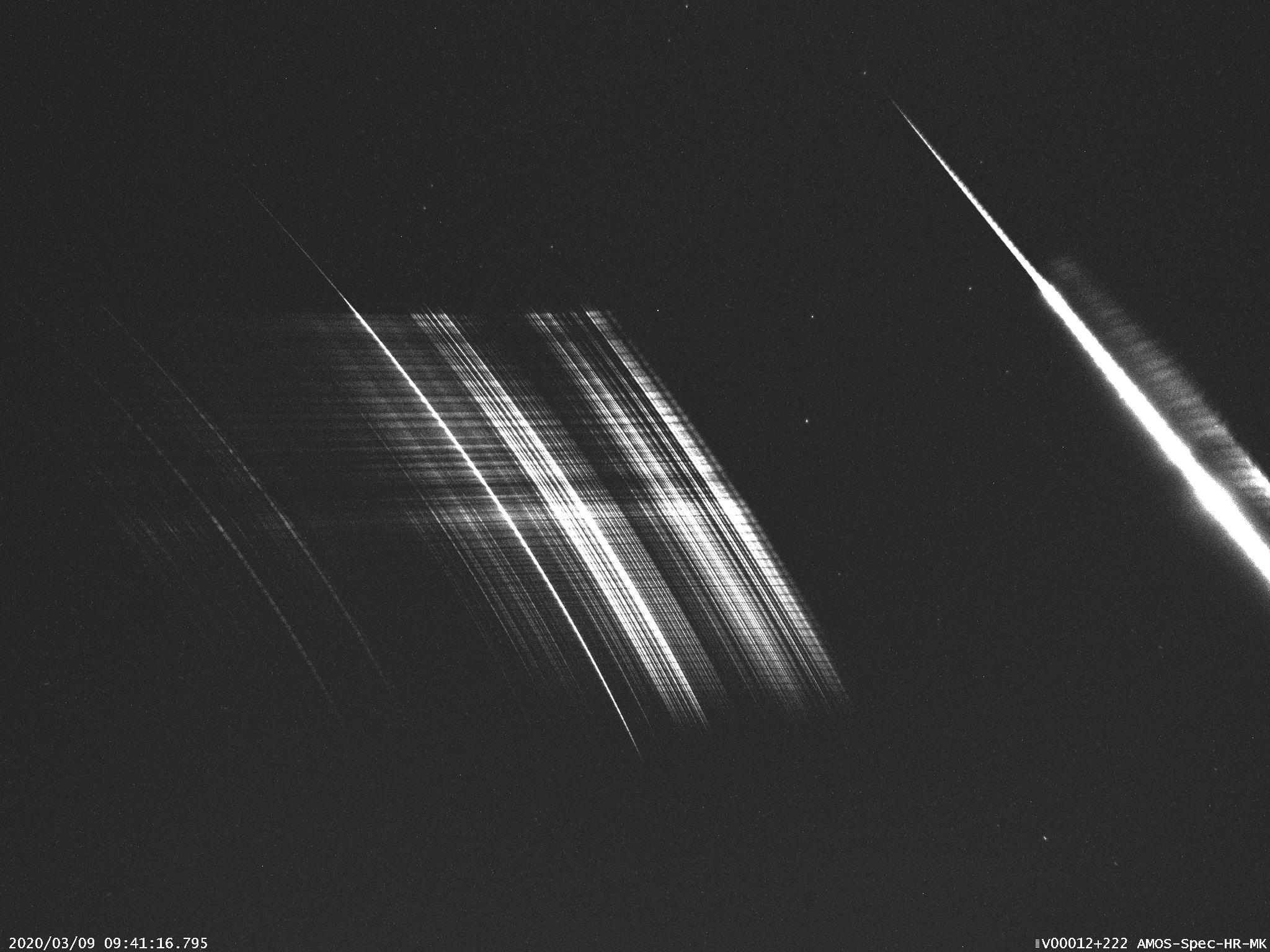}}
\caption{A composite image of meteor \small{M20200309$\_$094116} and its spectrum captured AMOS-Spec-HR station at the Manuakea Observatory in Hawaii.} 
\label{spectrum}
\end{figure}

The lower resolution spectrograph AMOS-Spec is based on a 30 mm f/3.5 fish-eye lens, an image intensifier, a projection lens, a digital camera and a holographic grating with 1000 grooves/mm. This setup provides a 100$^{\circ}$ circular FOV with a resolution of 1600x1200 pixels and frame rate of 12/s. The resulting spectra are captured at a dispersion of 1.3 nm/px (R $\approx$ 165). The higher resolution spectral systems AMOS-Spec-HR have smaller FOV of 60$^{\circ}$x45$^{\circ}$ and a resolution of 2048x1536 px resulting in a dispersion of 0.5 nm/px (R $\approx$ 500). The limiting magnitude of a meteor to be captured with a spectrum is approximately -1.5 mag for both systems. An example of a meteor spectrum recording by the AMOS-Spec-HR system is shown in Fig. \ref{spectrum}.

Emission spectra of ablating meteorites were obtained during simulated meteor experiments in the plasma wind tunnel facilities in cooperation between the Comenius University and the High Enthalpy Flow Diagnostics Group at the Institute of Space Systems, University of Stuttgart \citep{2017ApJ...837..112L}. The plasma flow condition used with local mass-specific enthalpy of 70\,MJ/kg at 
a stagnation pressure of $\sim$24\,hPa was set to recreate the ablation of meteoroids with a diameter of $\sim$4\,cm in the Earth's atmosphere at around 80 km altitude entering with a velocity of $\sim$12\,km/s. This flow condition was previously successfully tested for the characterization of the Hayabusa re-entry~\citep{Hermann_2016_07} and meteorite ablation \citep{2017ApJ...837..112L, 2018A&A...613A..54D}. Meanwhile, there are emission spectra of more than 20 different meteorite types of different composition available, many of them detected by both the high-resolution HEFDiG Echelle spectrograph and the meteor spectrograph AMOS-Spec-HR from the Comenius University.

The Echelle instrument of HEFDiG is a fiber-fed system designed for the wavelength interval 250 to 880\,nm. The pixel resolution varies over this interval from 43 to 143\,pm/px. An optical system was designed to capture the artificial meteor in the plasma wind tunnel in order to detect the spectral signature of the different materials. The frame rate of the system adjusts depending on the chosen exposure time, therefore, data was acquired with the highest camera gain possible and the exposure time was then reduced as short as a useful signal-to-noise ratio allows in order to have the frame rate as high as possible. The settings changed over the years of experimental campaigns, also because the system itself was improved for better signal capturing in 2019. For the comparison of data, this does not play a role. 


\subsection{Data processing}

The positions of meteors and reference stars were determined using our own detection program \textsc{AMOS-capture}. The astrometric reduction and heliocentric orbit determination was performed using our software \textsc{MT} (Meteor Trajectory), upgraded from an earlier version described in \citet{KornosIMC17}. The astrometric reduction is based on the all-sky procedure of \citet{1995A&AS..112..173B}. The program uses mathematical transformation for computing rectangular coordinates of catalog stars that are compared with measured stars. The identified stars are used to determine the 13 plate constants from which the zenith distance and azimuth of a meteor are computed. As a reference, the SKY2000 Master Catalog, Version 4 \citep{2001yCat.5109....0M} is used with stars up to +6th magnitude including the color indices. The precision of the AMOS astrometry (standard deviation of star positions) is approximately 0.02$^\circ$ – 0.03$^\circ$ for the current AMOS stations in Canary Islands, Chile, Hawaii and Australia, and approximately 0.03$^\circ$ – 0.05$^\circ$ for the stations in Slovakia. Overall, this translates to an accuracy of 10 - 100 meters for the atmospheric meteor trajectory.

The orbit determination method within \textsc{MT} is based on the standard plane intersection method of \citet{1987BAICz..38..222C}, with custom modules for the speed fit models, time-shift analysis and Monte Carlo simulations for error estimation. The meteor photometric reduction is also implemented in the \textsc{MT} software. It uses a brightness calibration based on reference stars for fainter meteors (-2 mag and fainter), and a saturation correction for brighter fireballs, based on a comparison with bright planets and Moon in different phases. Each meteor analyzed in this work was processed and analyzed manually. The details of the AMOS data processing pipeline are in preparation for publication (Tóth et al., in preparation).

The spectral reduction and analysis were performed following the procedures described in \citet{2019A&A...629A..71M}. Each analyzed spectrum was manually scanned in individual frames of the meteor video recording, calibrated and fitted with a synthetic spectrum comprised of the main atomic and molecular emission features of the meteor. The synthetic spectrum was fitted to the calibrated meteor spectrum using the damped least-squares method (the Levenberg – Marquardt algorithm) within the \textsc{Fityk} software \citep{Wojdyr:ko5121}. Gaussian instrumental line profiles were assumed in the synthetic spectra, with characteristic FWHM typically 1.2 nm for AMOS-Spec-HR data, and 3 nm for AMOS-Spec data. The uncertainties of the measured line intensities were calculated from their signal to noise ratio (SNR) in each spectrum. The multiplet numbers used here are taken from \citet{1945CoPri..20....1M}.

\begin{figure}
\centerline{\includegraphics[width=.95\columnwidth,angle=0]{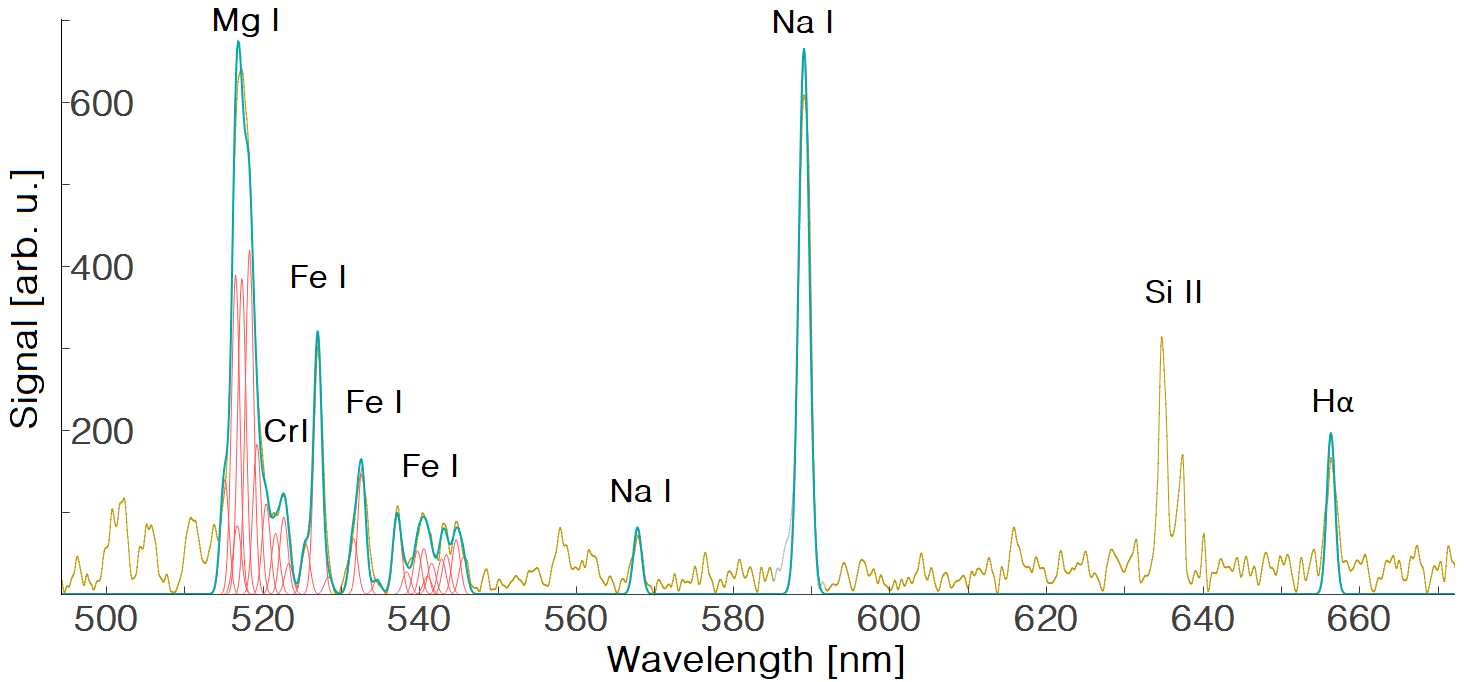}}
\centerline{\includegraphics[width=.95\columnwidth,angle=0]{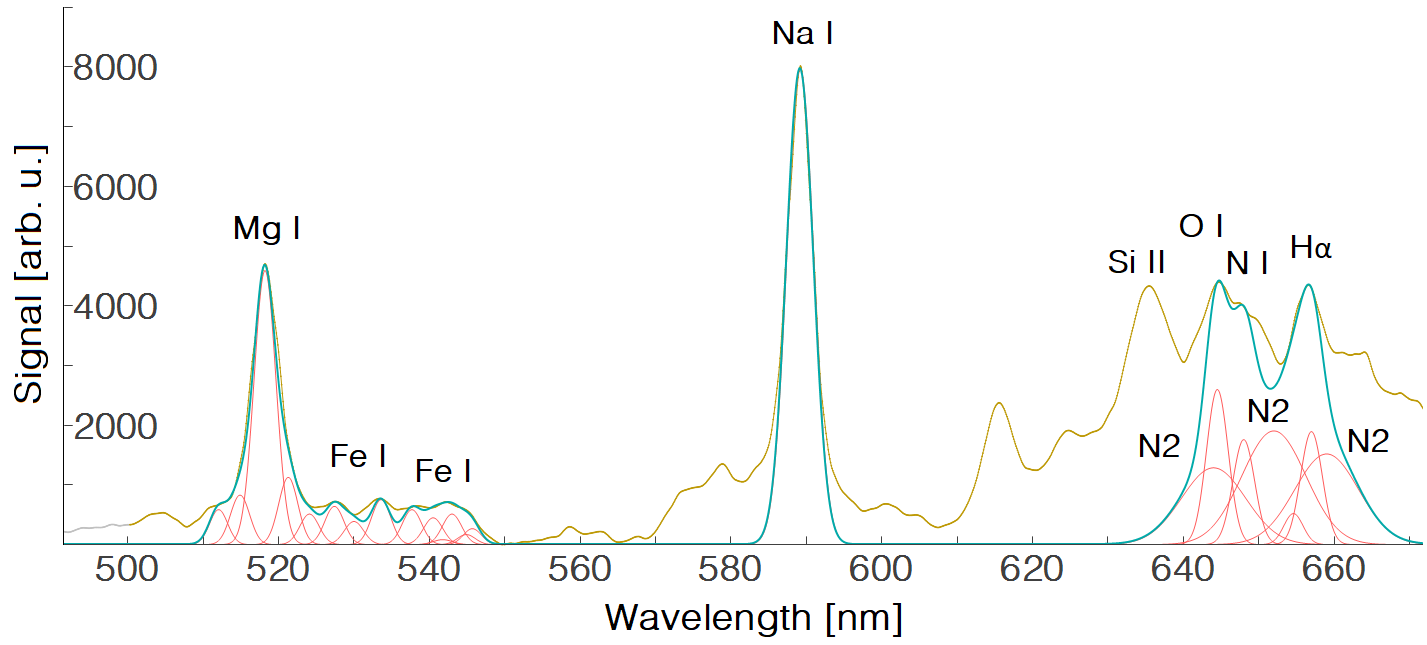}}
\caption{Examples of the fit of the synthetic spectra (blue-green) on measured meteor spectra (dark yellow) as a convolution of the main emission contributions (red) in the 500-665 nm region. The fit of higher resolution data of a mu-Leonid observed by AMOS-Spec-HR (upper panel) is compared with a fit of lower resolution data of a Perseid observed by AMOS-Spec (lower panel). The main measured emission lines and bands are marked.} 
\label{fit}
\end{figure}

In this work, we focus in detail on the detection and relative intensity of the H$\alpha$ line positioned near 656.3 nm. When present in meteor spectra, this line is typically the strongest feature in the 650-660 nm range, making its measurement usually straightforward in higher resolution spectra. In lower resolution spectra, we also fitted the contributions from near emission lines of the atmospheric O I (O I-9 multiplet near 645.5 nm), N I (N I-21 multiplet near 648.3 nm) and strongest N$_2$ first positive $B^{3}\Pi_{g}-A^{3}\Sigma u^{+}$ bands (near 644, 652 and 659 nm), as well as the Fe I lines (Fe I-268 multiplet near 654.6 nm) originating in the meteoroid, all of which can affect the intensity of H$\alpha$. The intensities of these contributions were individually adjusted, as they are dependent on the meteor speed (temperature of the radiating plasma) and the estimated iron content in each meteoroid. The contribution of the nearby Ca I line (Ca I-1 multiplet near 657.2 nm), noted as a possible influence by \citet{2004AsBio...4..123J}, was not recognized as a significant contribution in most meteor or laboratory ablation data. The comparison of the meteor spectrum fit in lower resolution and higher resolution data are displayed in Fig. \ref{fit}. The emission of H$\alpha$, CN and other compared multiplets (mainly Fe I-15 lines around 526-550 nm, Mg I-2 triplet at 516-519 nm and N I-21 lines at 648-649 nm) in the Echelle spectra of ablated meteorites were measured by the same procedure. Given the high-resolution (R $\approx$ 5000) of the Echelle data, the contributions of the surrounding lines to H$\alpha$ were negligible, and resulted in higher accuracy of the analyzed intensity ratios.

The original heliocentric orbits of the analyzed meteoroids were classified using the Tisserand parameter with respect to Jupiter $T_J$, which distinguishes between orbits characteristic for asteroidal bodies (AST) from the main-belt or NEO space ($T_J>$ 3), Jupiter-family (JF) type cometary bodies from the Kuiper belt (2 $<T_J<$ 3 and i $<$ 45°) and Halley-type (HT) cometary bodies from the Oort cloud ($T_J<$ 2 and i $>$ 45°) \citep{1979aste.book..289K}. 

The meteoroid material strengths were estimated based on the empirical fireball classification parameter $P_E$ \citep{1976JGR....81.6257C}. $P_E$ is derived from the meteor terminal heights, entry speed, radiant zenith distance and photometric mass, and can be used to roughly differentiate between stronger asteroidal materials similar to ordinary chondrites (Type I), carbonaceous type meteoroids (Type II), standard (Type IIIA) and fragile cometary (Type IIIB) meteoroids.

The Echelle emission spectra of ablating meteorites captured during the laboratory wind tunnel testing were calibrated after each actual experiment by positioning a calibration lamp at the position of interest in the facility. The known radiance of the calibration lamp is then used to translate the cameras ADU to the measured radiance of the meteorite sample. For a typical meteorite experiment with a duration of $\sim$4\,s of duration, between 15 and 70 frames are acquired. The final spectra of each meteorite used for line intensity measurements performed in this work were obtained by summing all calibrated frames of the spectra captured during actual ablation of the meteorite and subtracting the spectra baselines. Line intensities were measured using the Fityk program, similarly to AMOS data.

\section{Results} \label{sec:results}

\subsection{Hydrogen emission in meteors from observational data} \label{sec:survey}

We have conducted a survey of 304 meteor spectra produced by mm- to dm- sized meteoroids (corresponding to meteors of -1 to -14 abs. mag), identifying 62 meteors with confirmed H$\alpha$ emission. These samples formed the basis for further analysis. The entry speeds and trajectory parameters were determined for 52 meteoroids, the orbital elements (excluding inaccurate solutions) were determined for 40 meteoroids and material strength properties for 39 of these meteoroids.

\begin{figure}
\centerline{\includegraphics[width=.9\columnwidth,angle=0]{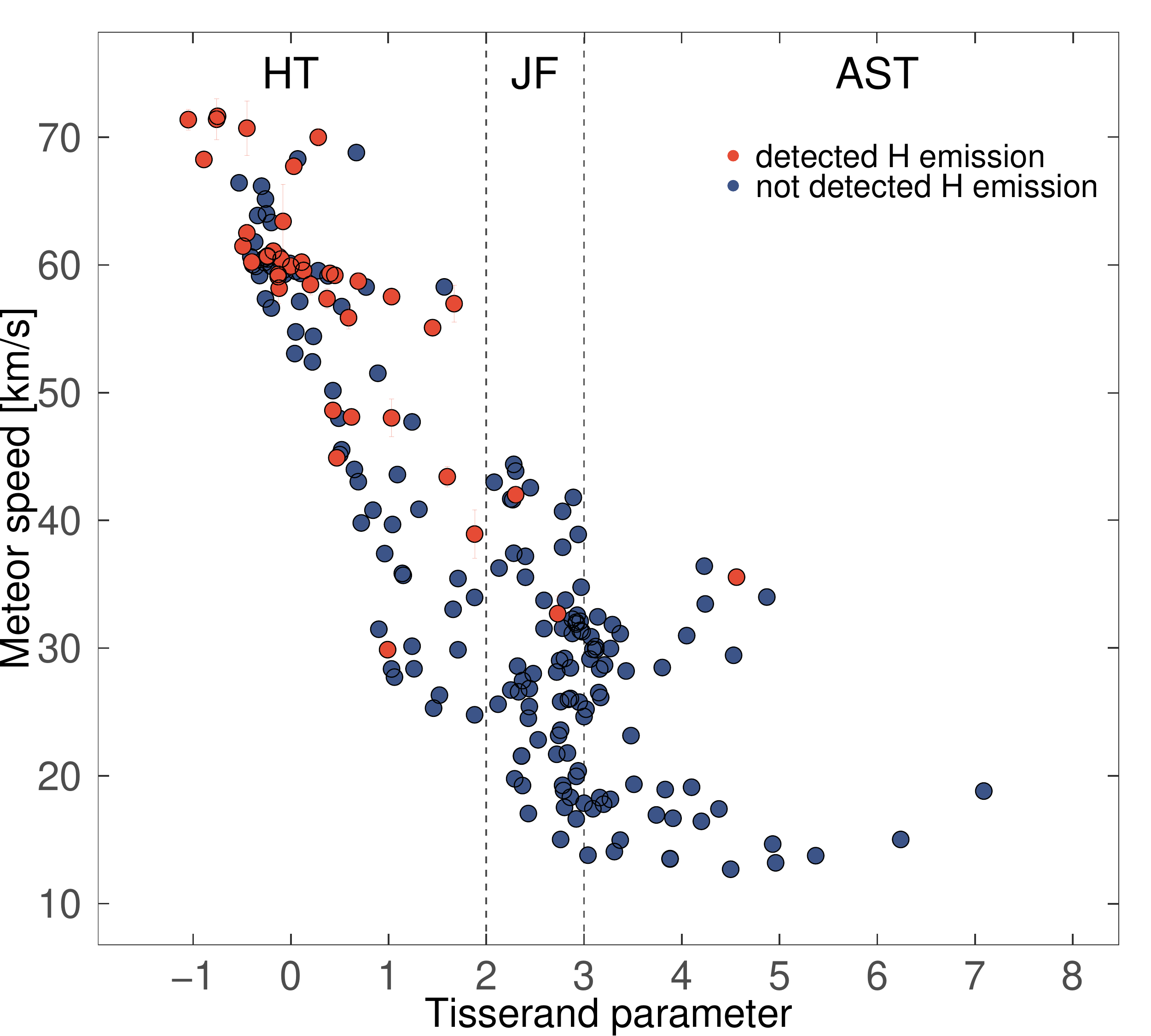}}
\caption{Entry speeds as a function of the Jovian Tisserand parameter ($T_J$) for the analyzed meteoroids. The plot displays the orbital sources of meteoroids with and without the detected H$\alpha$ emission in their spectra. The $T_J$ classification is divided into three characteristic zones: Halley-type cometary orbits -- HT, Jupiter-family type cometary orbits -- JF and asteroidal orbits -- AST.} 
\label{vi_Tj_Ha_nonHa}
\end{figure}

It is expected that increased H emission will be characteristic for cometary meteoroids, due to their higher content of organic compounds and water molecules (i.e., as ice and bound in minerals) compared to asteroidal meteoroids. This assumption is clearly demonstrated in our data, which shows that $\approx$ 92\% of meteoroids with recognized H emission originate from longer-period Halley-type cometary orbits (Fig. \ref{vi_Tj_Ha_nonHa}). The median entry speed of meteors with detected H emission is around 59 km s\textsuperscript{-1}, partly due to the large contribution of Perseid meteors ($v_i \approx$ 60 km s\textsuperscript{-1}) in our sample.

As noted in Section \ref{sec:intro}, due to its high excitation, stronger H$\alpha$ line is favored in faster and brighter meteors (which produce the high-temperature spectral component). High meteor speeds are in general characteristic for meteoroids on cometary orbits. It can be assumed that two factors contribute to the detection of H emission in meteors: sufficient temperature of the radiating plasma (presented in observational data by meteor speed) for the H$\alpha$ line to be visible, and meteoroid composition with sufficient content of H-bearing compounds. 

\begin{figure}
\centerline{\includegraphics[width=.9\columnwidth,angle=0]{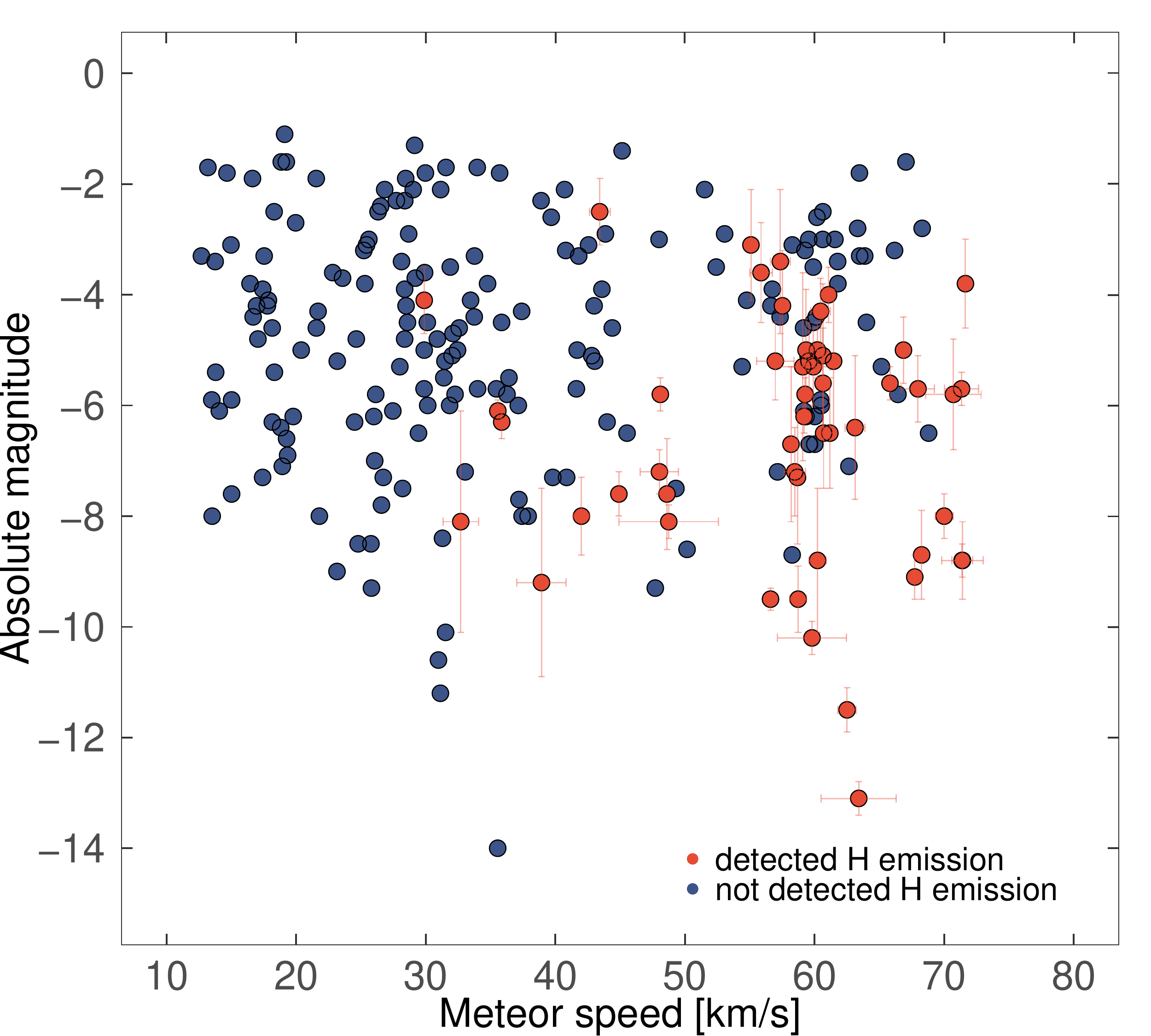}}
\caption{The absolute magnitudes and entry speeds of meteors with and without the detected H$\alpha$ emission. The plotted meteors were caused by meteoroids of mm to dm size. Errors bars are indicated only for meteors with detected H emission.} 
\label{absMag_vi_Ha_nonHa}
\end{figure}

Fig. \ref{absMag_vi_Ha_nonHa} indeed shows that slow meteors ($v_i <$ 30 km s\textsuperscript{-1}) do not exhibit H$\alpha$ emission. However, we did not find clear continuous dependency between meteor speed and H$\alpha$ line intensity once the lower limit speed ($v_i \approx$ 30 km s\textsuperscript{-1}) was achieved. We note that a continuous dependence between the line intensity and meteor speed can be observed for other meteoric elements, most notably the volatile Na, which is favored in low velocity meteors \citep{2005Icar..174...15B, 2020Icar..34713817M}.

Further in this section, we therefore only assess the presence and relative intensity of H$\alpha$ in faster cometary meteors, without interpreting the lack of H emission detected in slower, often asteroidal bodies. Hydrogen emission from asteroidal materials is later discussed in Section \ref{sec:laboratory} based on laboratory data of ablated meteorite samples.

\begin{figure}
\centerline{\includegraphics[width=.90\columnwidth,angle=0]{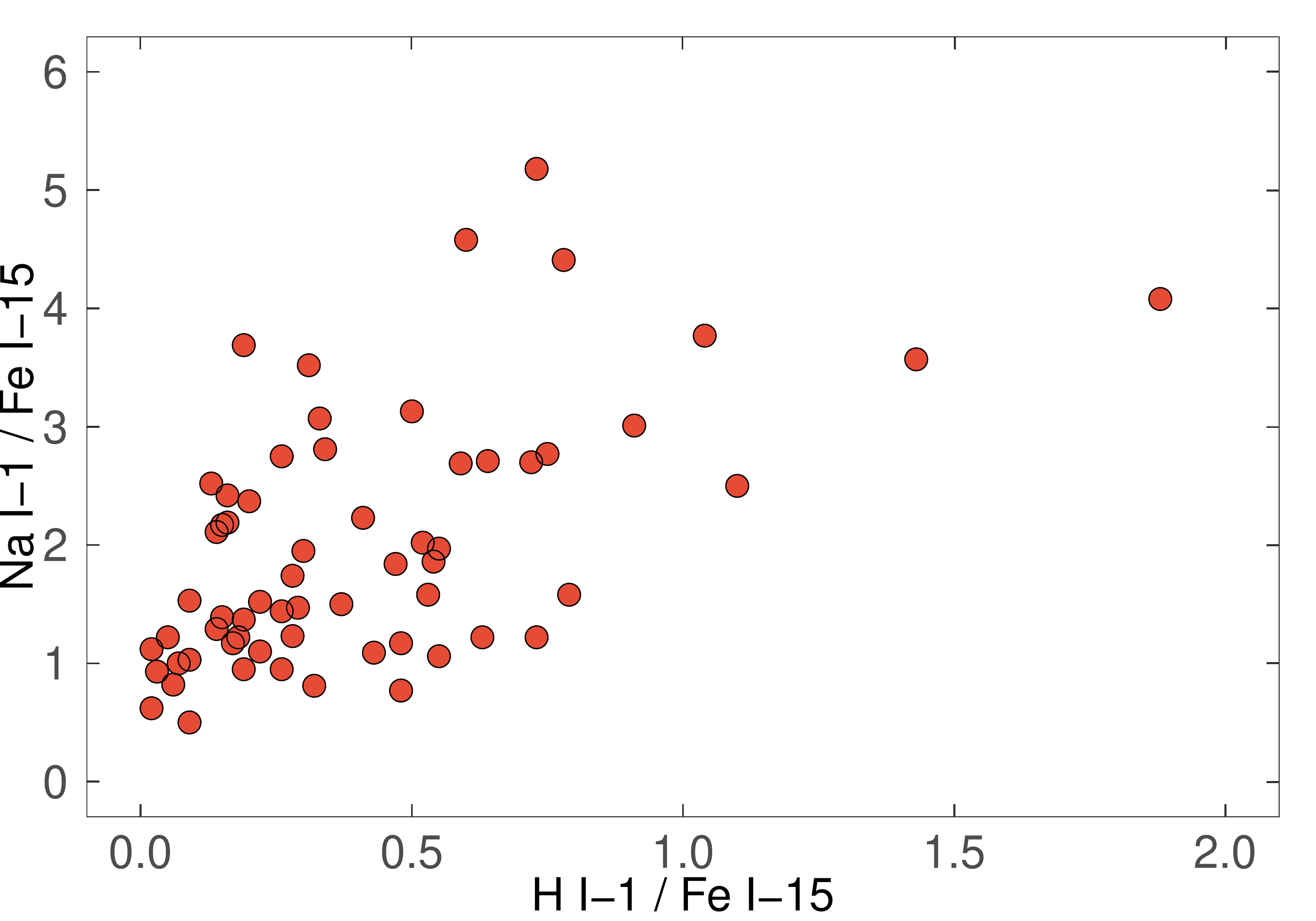}}
\caption{Observed Na I-1/Fe I-15 intensity ratio as a function of H$\alpha$/Fe I-15 intensity ratio based on the spectral data of 48 meteors with the detected H$\alpha$ line.} 
\label{HaFe_vs_NaFe}
\end{figure}

Since the H emission in meteors is associated with the presence of volatile ices and/or hydrated minerals, and the most commonly detected tracer of volatiles in most meteor spectra is sodium, we have searched for a correlation between the H I and Na I line intensities in meteors. The results depicted in Fig. \ref{HaFe_vs_NaFe} suggest that, on average, meteoroids with stronger Na line also exhibit higher H I intensities. Both Na I and H I lines can be used to trace volatile content in meteoroids, but one must carefully consider the effects of meteor speed which can affect intensities of the two lines differently.

\begin{figure}
\centerline{\includegraphics[width=.90\columnwidth,angle=0]{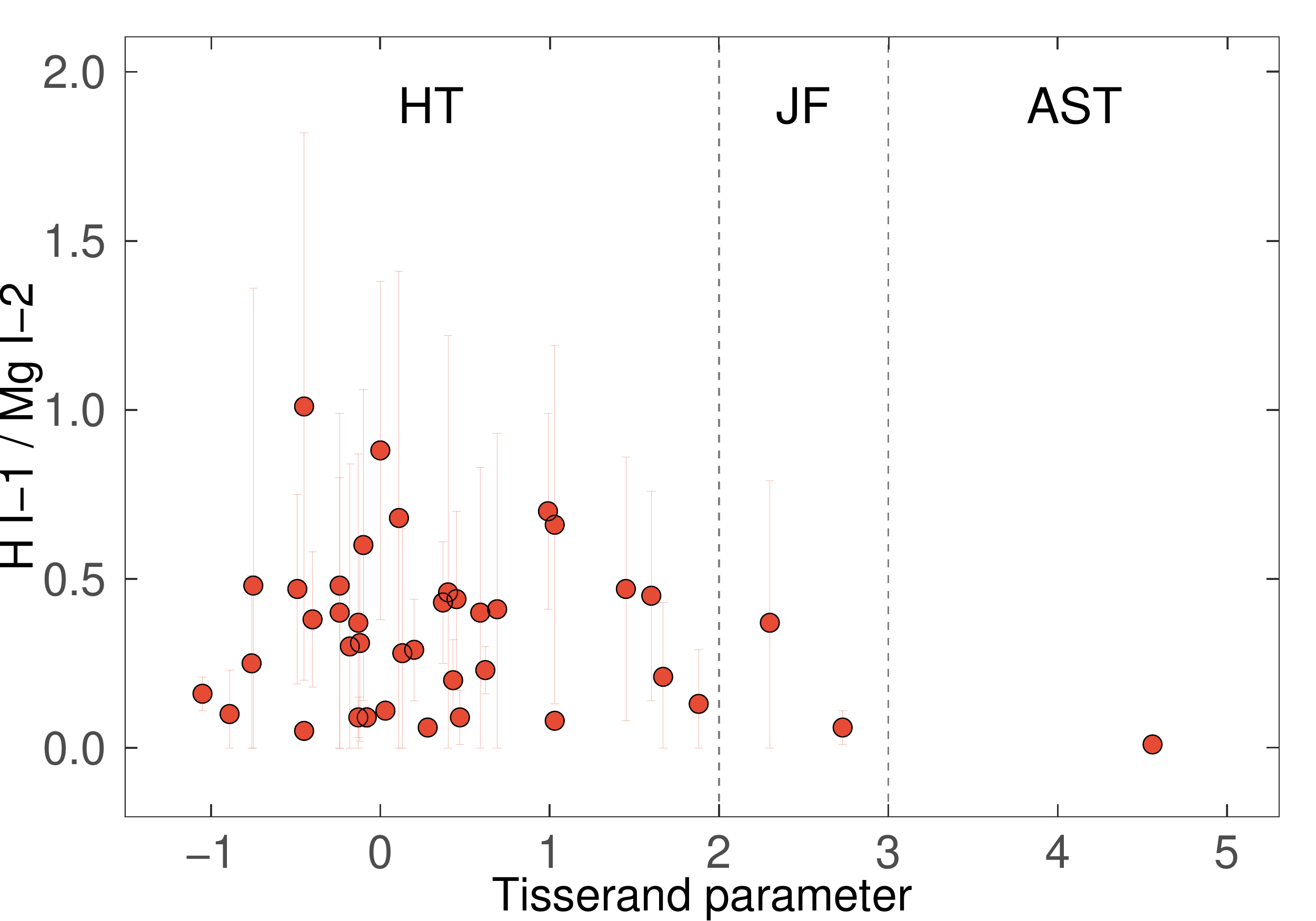}}
\caption{The dependence of the H$\alpha$/MgI-2 line intensity ratio on the Jovian Tisserand parameter.} 
\label{HaMg_vs_Tj}
\end{figure}

\subsubsection{Halley-type orbits}

As shown in Fig. \ref{vi_Tj_Ha_nonHa}, the vast majority of meteoroids with detected H emission originated on Halley-type and long-period orbits. Our results also suggest that the measured H/Mg intensity ratio is on average increased in meteoroids on orbits with lower Jovian Tisserand parameter (Fig. \ref{HaMg_vs_Tj}), though the data for meteoroids from Jupiter-family type and asteroidal orbits is quite limited. This effect might reflect the fact that meteoroids on longer-period cometary orbits are on average less affected by space weathering effects (most notably the solar radiation) and thus retain more hydrogen bound in ices and/or hydrated minerals and organic compounds.

We have found that the H emission can be detected in most fast ($>$ 50 km s\textsuperscript{-1}) and bright ($<$ -4 mag) fireballs (Fig. \ref{absMag_vi_Ha_nonHa}), typically caused by cometary meteoroids from Halley-type orbits. However, the then open question is, why some bright fireballs from cometary orbits (Fig. \ref{vi_Tj_Ha_nonHa}) did not exhibit any H emission. For fainter ($>$ -4 mag) and slower ($v_i <$ 40 km s\textsuperscript{-1}) meteors, the detection of H$\alpha$ can be hampered by the occasionally low recorded S/N of the spectrum near the 600 - 700 nm range, depending on the specific atmospheric conditions and the meteoroid composition. We have performed individual inspection of the remaining 35 events brighter than -4 mag and faster than 40 km s\textsuperscript{-1} with no detected H emission. For many of these cases, the background noise level around the H$\alpha$ wavelength, caused by a combination of effects including strong blending of surrounding N I and N\textsubscript{2}, diffraction geometry and atmospheric conditions, resulted in S/N $<$ 2 at the H$\alpha$ position. We note that the lower H$\alpha$ S/N in these cases was potentially also in part affected by the lower compositional H content of the meteoroid. The detection of H$\alpha$ was in these cases therefore considered unreliable. Only in the remaining 9 fast and bright cometary meteors were both atmospheric and observational conditions very favorable, but no H emission was detected. We assume that in these cases, most of the hydrogen was lost from the meteoroid through space weathering processes, dominantly through thermal desorption induced by solar radiation during close perihelion approaches (4 of the remaining events have q $<$ 0.25 au). Another process potentially causing volatile loss in meteoroids, which do not have low perihelia, is related with cosmic ray irradiation. It is predicted that long-term cosmic ray irradiation of cometary surfaces in the Oort cloud may lead to a formation of volatile-free refractory crust \citep{1992A&A...266..434S, 2005Icar..174...15B}. The disruption of the crust \citep{2002Sci...296.2212L} can then produce meteoroids on longer-period orbits that are depleted in volatile elements like Na or H.

\begin{figure}
\centerline{\includegraphics[width=.80\columnwidth,angle=0]{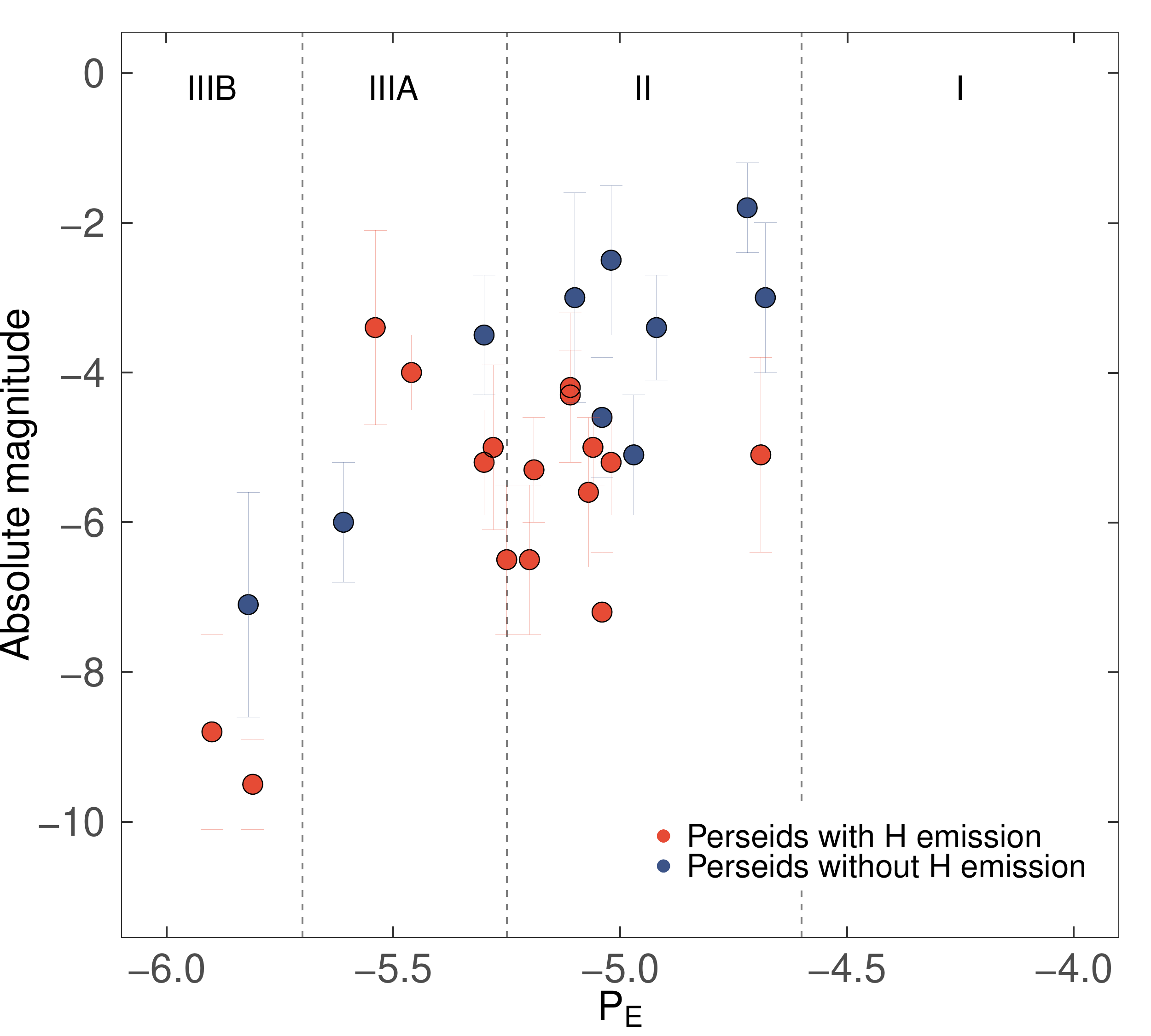}}
\caption{The detection of H$\alpha$ emission in Perseid meteoroids plotted as functions of their absolute magnitude and $P_{E}$ coefficient of meteoroid strength. The detection of H$\alpha$ in meteors $>$ -4 mag may be hampered by the lower S/N of the spectrum. The undetected H emission in four brighter Perseids appears to be related with the observational conditions: clouds, moonlight and acute meteor entry angle, also resulting in lower S/N at the H$\alpha$ position.} 
\label{PER1}
\end{figure}

\begin{figure}
\centerline{\includegraphics[width=.80\columnwidth,angle=0]{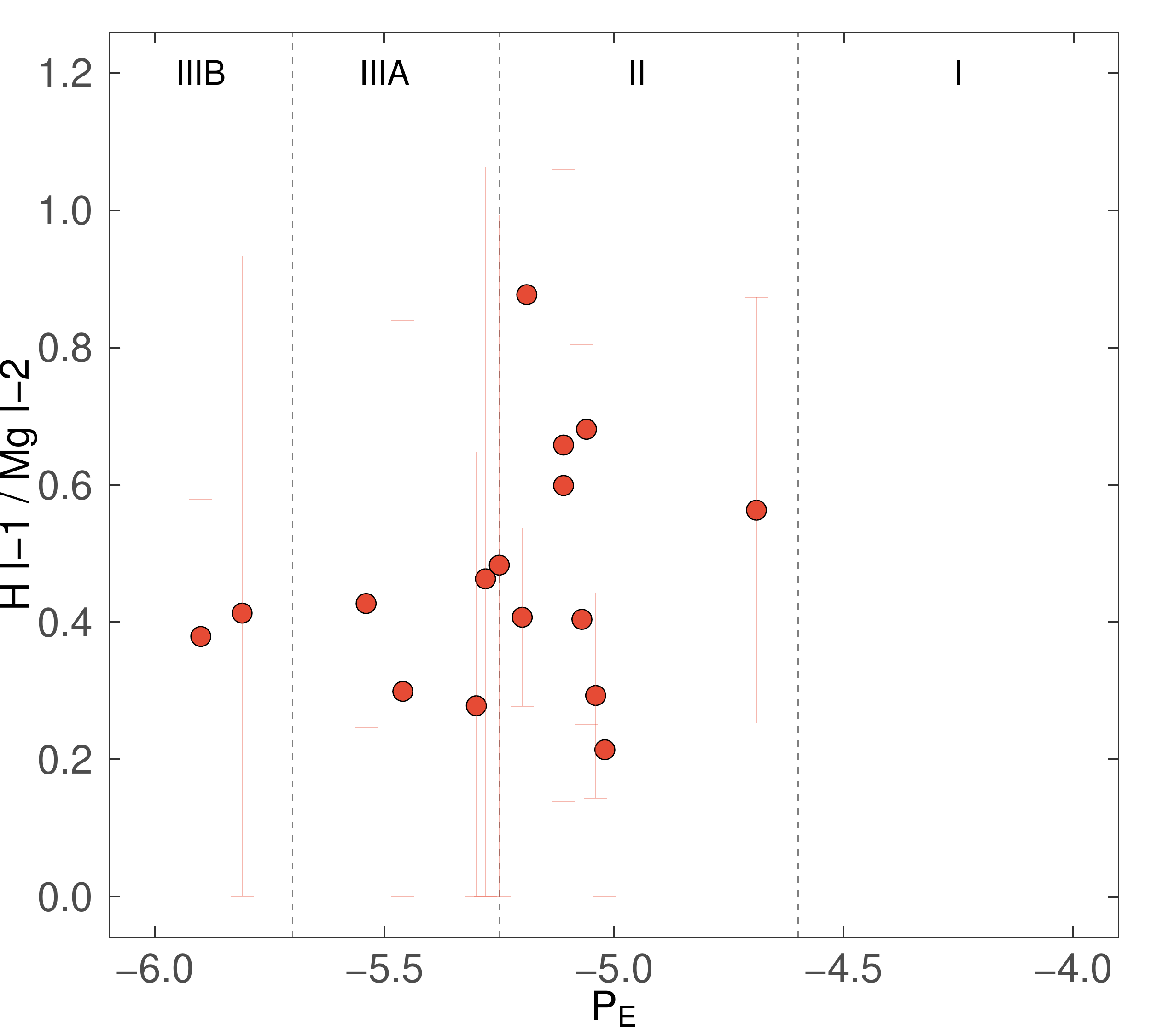}}
\caption{The H$\alpha$/MgI-2 line intensity ratio as the function of the $P_{E}$ coefficient for Perseid meteoroids.} 
\label{PER2}
\end{figure}

As a case study of a Halley-type stream, we may look at the best represented shower in our dataset - Perseids with 26 analyzed meteor spectra. Since all Perseids have similar speed, their ablation temperatures may only differ, to a lesser degree, with meteoroid size and individual physical properties such as porosity and related bulk strength. The results in Fig. \ref{PER1} confirm that most Perseids with no detected H emission were fainter ($>$ -4 mag), resulting in lower S/N of the spectrum. The lack of detected H emission in four brighter Perseids appears to be also affected by unfavorable observational conditions and blending with surrounding N I and N\textsubscript{2} emission which resulted in low S/N at the H$\alpha$ position. We did not find clear correlation between the relative H$\alpha$ intensity ratio and material strength defined by the $P_{E}$ coefficient (Fig. \ref{PER2}), though it can be noted that the four Perseids with H$\alpha$/MgI-2 $>$ 0.6 were Type II fireballs. These results also point out relatively large variations of H/Mg intensity ratios within Perseids, suggesting significant heterogeneity of the present H-bearing compounds content within the debris from comet 109P/Swift–Tuttle. We note that a relatively high degree of heterogeneity in elemental intensity ratios (e.g. Fe/Mg and Na/Mg) among meteoroids originating from one parent body appears to be common \citep{2015A&A...580A..67V, 2017P&SS..143..104M, 2020P&SS..19405040A}, resulting from environmental factors and thermal history, but probably also reflects real compositional inhomogeneities within comet on mm-dm scales.

Besides Perseids, the dataset of meteoroids with detected H emission originating from Halley-type cometary orbits included four Leonids from 55P/Tempel–Tuttle, two Lyrids from C/1861 G1 (Thatcher), two zeta Cassiopeiids, one Orionid from 1P/Halley, one sigma Hydrid, one 49 Andromedid and individual samples from few other minor streams (December omega Ursae Majorids, 58 Piscids, mu Leonids and 6 Sextantids) from the IAU Meteor Data Center Working List \citep{2017P&SS..143....3J}.

\subsubsection{Jupiter-family and asteroidal orbits: volatile loss}

\begin{figure}
\centerline{\includegraphics[width=.95\columnwidth,angle=0]{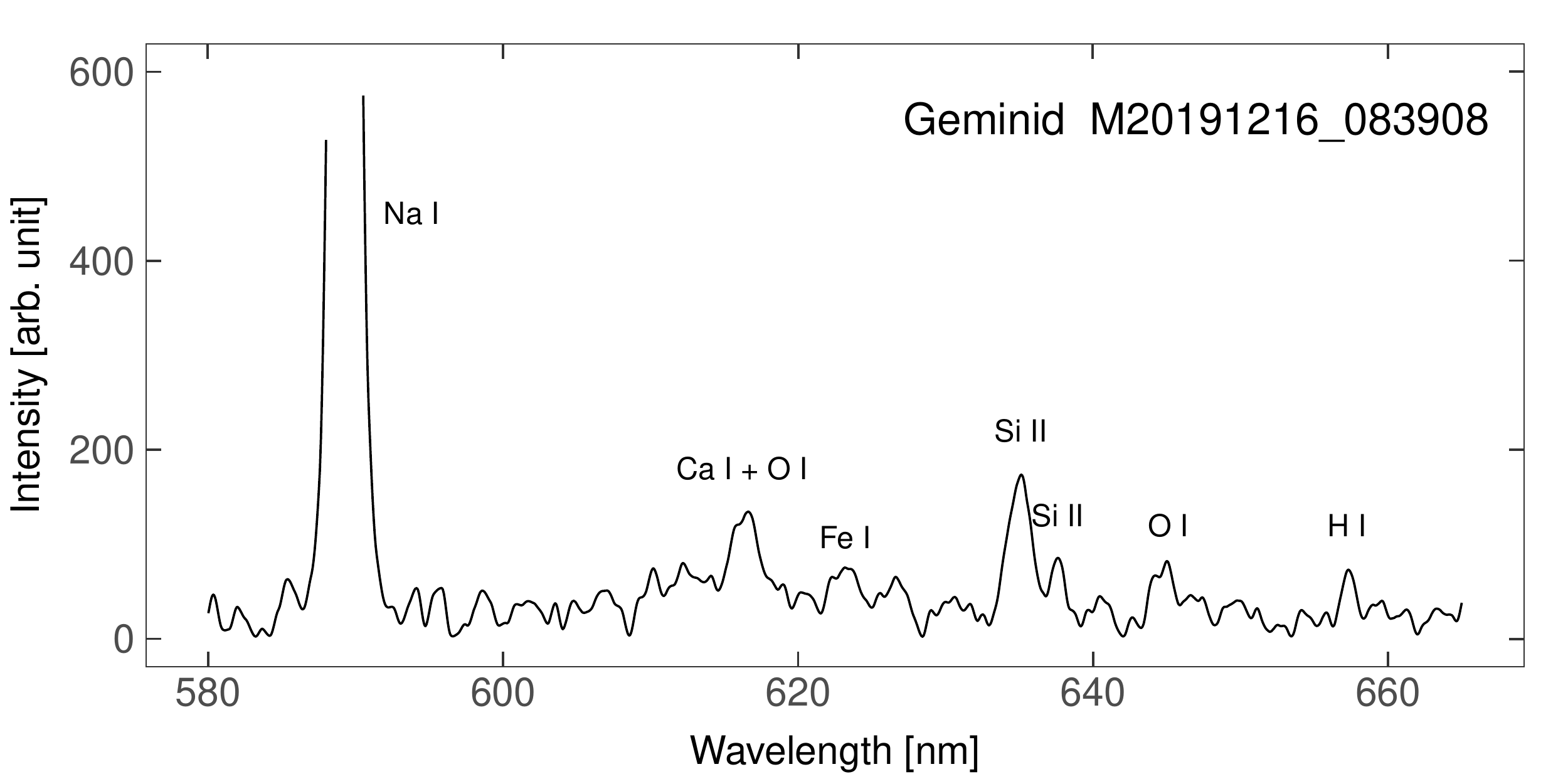}}
\centerline{\includegraphics[width=.95\columnwidth,angle=0]{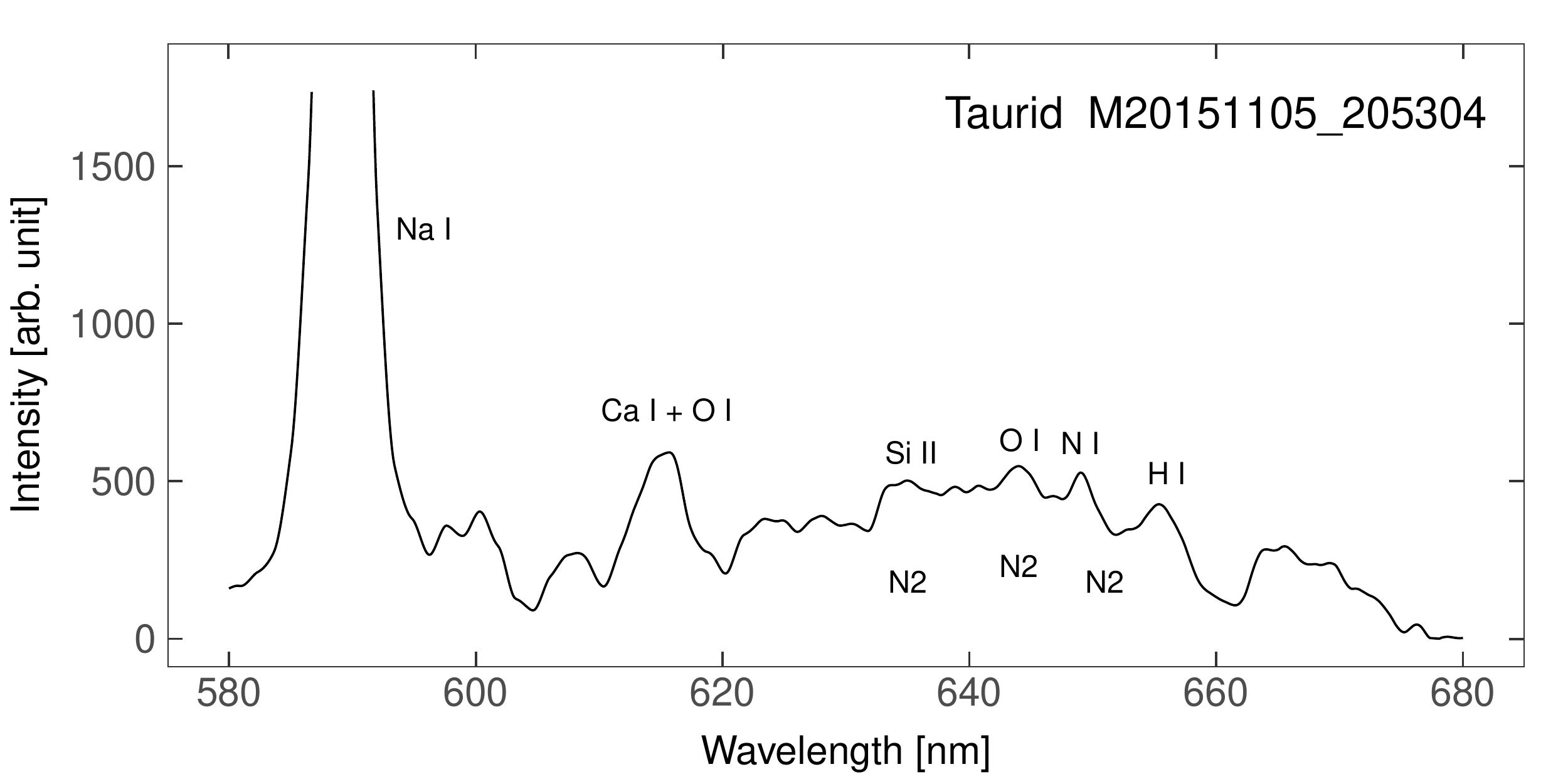}}
\caption{The emission spectra of a Geminid and a northern Taurid meteoroid with detected hydrogen emission in the 580-680 nm region. The meteoroid code designates the UTC time of detection in YYYYMMDD\_HHMMSS format. The peak of the Na I-1 doublet is outside the plot limits.}
\label{GeminidTauridspec}
\end{figure}

We have found only one meteoroid with recognized H emission that originated from an asteroidal orbit ($T_J>$ 3). This meteoroid (Fig. \ref{GeminidTauridspec}) belongs to the Geminid meteoroid stream originating in (3200) Phaethon. As evidenced by the observed activity of (3200) Phaethon \citep{2010AJ....140.1519J}, Geminids cannot be considered as a typical asteroidal material. The spectral properties and the suspected origin of (3200) Phaethon in the carbonaceous (2) Pallas with aqueously altered surface \citep{2010A&A...513A..26D, 2020NatAs...4..569M} give credibility to the possible presence of organic matter in Geminids. 

The detection of hydrogen in a Geminid raises interesting questions about the preservation of volatiles embedded in ices or hydrated minerals producing H emission. Geminids have characteristically close perihelion approaches to the Sun (q $\approx$ 0.14 au), at which solar heating can cause thermal metamorphism and dehydration of the meteoroid. The loss of volatiles is evidenced by the directly observed depletion of sodium in Geminid meteoroids \citep{2005Icar..174...15B, 2019A&A...629A..71M, 2020P&SS..19405040A}. Variations of Na intensity are often observed among Geminids, thought to be related with the age of the meteoroid as a separate body (i.e. the time from its release from the parent body). Meteoroids which suffered fewer perihelion passages are thought to retain more volatiles. The rate of the volatile loss also seems to be related with the size of meteoroids, since smaller bodies show signs of more significant sodium depletion \citep{2019A&A...629A..71M, 2020P&SS..19405040A}. The Geminid with detected H emission also shows strong Na I-1 lines, suggesting that the meteoroid was released from its parent relatively recently and likely did not suffer strong solar radiation, which enabled the preservation of volatiles. Another Geminid with detected H emission was previously found by \citet{2004AsBio...4..123J}. Our dataset included four more Geminids with no H emission. It appears that Geminids that did not suffer strong solar radiation may exhibit increased volatile presence through both Na and H. This effect should be further studied in a larger dataset of Geminid spectra.

\citet{2009PASJ...61.1375O} have found that the solar-radiation heating on Phaethon is a function of the latitude and thus its northern hemisphere may be more dehydrated (although a non-preferential heating of Phaethon was later implied by \citep{2016A&A...592A..34H}). Furthermore, recent studies suggest that Phaethon was originally hydrated and has since lost volatiles \citep{2020NatCo..11.2050T} on its surface. Our detection of hydrogen in a presumably recently separated Geminid meteoroid however gives validity to the hypothesis that some regions of the Phaethon surface may still be hydrated. 

\begin{figure}
\centerline{\includegraphics[width=.90\columnwidth,angle=0]{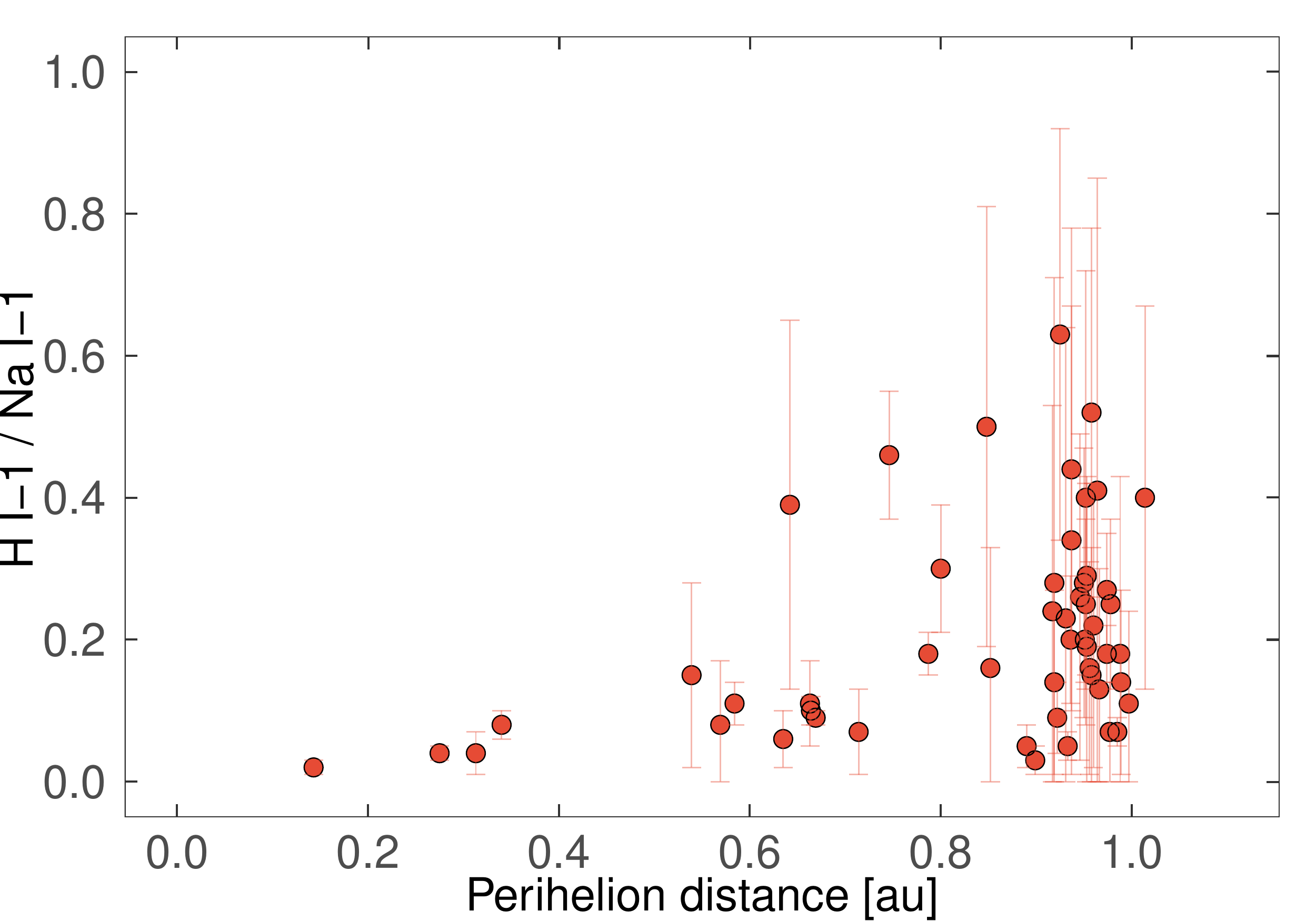}}
\caption{The observed H$\alpha$/Na I-1 intensity ratio as a function of perihelion distance for meteors with detected H$\alpha$ emission.} 
\label{HaNa_vs_q}
\end{figure}

Fig. \ref{HaNa_vs_q} depicts the dependency between H$\alpha$ intensity relative to Na I and perihelion distance for all meteoroids with detected H. All meteoroids with $q <$ 0.6 au have H$\alpha$/Na I-1 $<$ 0.2, while other meteoroids may exhibit increased H$\alpha$/Na I-1 ratios. Note that H$\alpha$ intensity is measured relatively to the lines of also volatile Na, which was previously found to be depleted in meteoroids on shorter perihelion orbits, relative to silicate elements such as Mg \citep{2019A&A...629A..71M}. This emphasizes the volatility of hydrogen in meteoroids, which seems to deplete even more efficiently than Na.

Hydrogen was detected only in two meteoroids on Jupiter-family type orbits (2 $<T_J<$ 3). These meteoroids include a northern Taurid (Fig. \ref{GeminidTauridspec}) from comet 2P/Encke and one sporadic meteoroid. A possible H emission was also analyzed in a southern $\delta$ Aquariid (SDA) originating from the sun-grazing comet 96P/Machholz, but due to low S/N level, its detection was not confirmed. A severe volatile loss is expected in the sun-grazing SDA meteoroids (q $\approx$ 0.07 au), but volatiles can be detected in the recently released or relatively preserved larger meteoroids. Variations of the volatile Na in SDA meteoroids was previously noted by \citet{2016P&SS..123...25R} and \citet{2019A&A...629A..71M}. 

In general, the Taurid stream exhibits high heterogeneity of material strengths and composition, and was speculated to include stronger asteroidal material \citep{2013M&PS...48..270B, 2016MNRAS.461..674O, 2017A&A...605A..68S}. Recent studies have however confirmed that the majority of Taurids have cometary characteristics and are likely produced by comet 2P/Encke \citep{2017P&SS..143..104M, 2020P&SS..18204849B}. Given its water production rate \citep{2001Icar..152..268M} and relatively close perihelion approaches, 2P/Encke is considered an ideal candidate for detection of perihelion-induced aqueous alteration on its surface \citep{2020Icar..35113956S}. The detection of H emission in a Taurid meteor shows that Taurid meteoroids may retain water and/or other H-bearing compounds. The studied sample included 18 more Taurids in which H emission was not detected. This fact is probably related with the generally lower ablation temperatures of Taurids at $v_i \approx$ 30 km s\textsuperscript{-1}, which make the detection of H$\alpha$ less probable (Fig. \ref{absMag_vi_Ha_nonHa}), but it could also point out the volatile depletion within the stream. 

\subsubsection{Material strengths}

\begin{figure}
\centerline{\includegraphics[width=.90\columnwidth,angle=0]{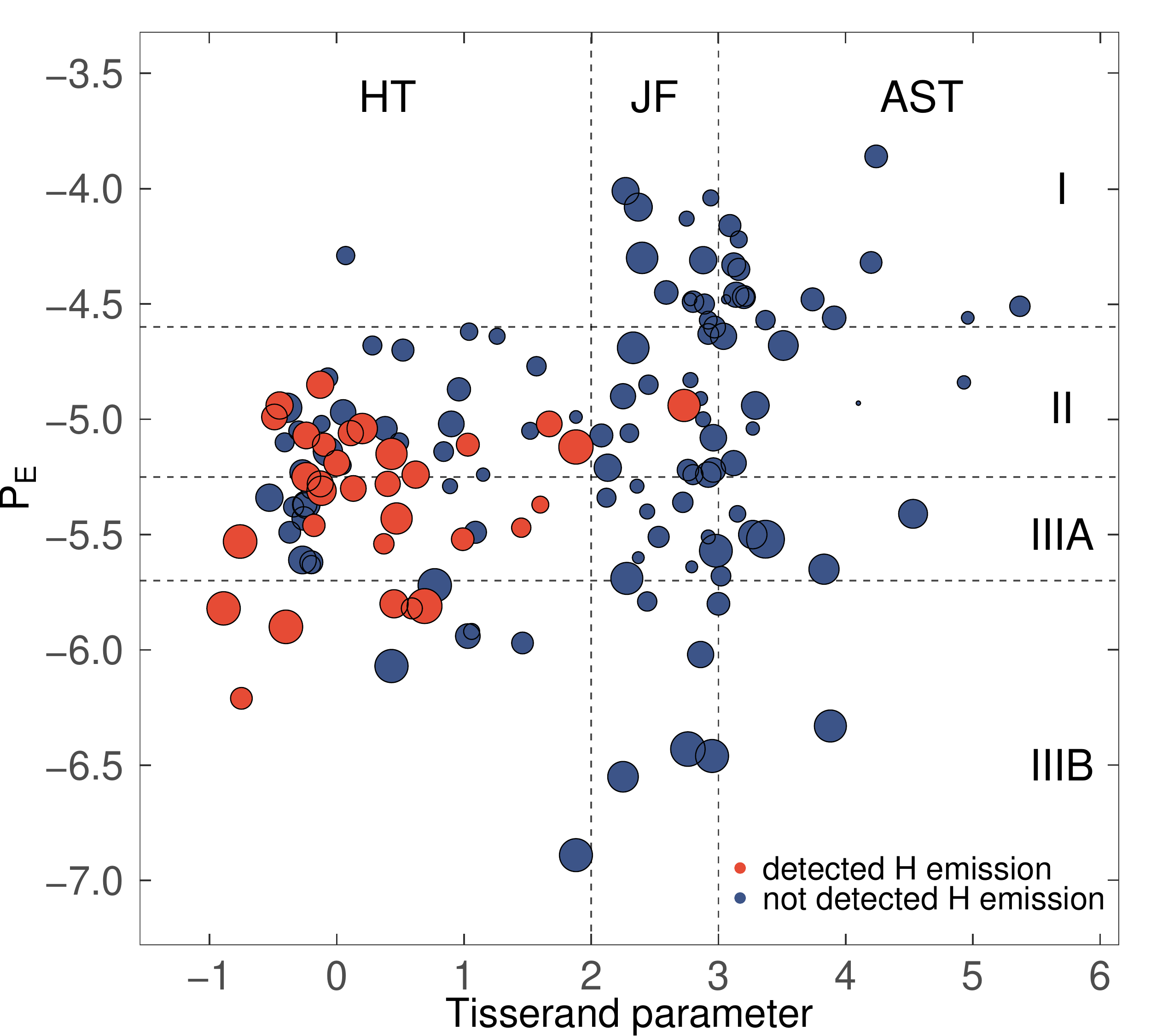}}
\caption{The dependency between the orbital classification of meteoroids based on the Jovian Tisserand parameter and material strength classification based on the $P_E$ criterion for all meteoroids in our sample. The size of the meteoroid symbols reflects the relative meteor magnitude.} 
\label{TJPE}
\end{figure}

\begin{figure}
\centerline{\includegraphics[width=.90\columnwidth,angle=0]{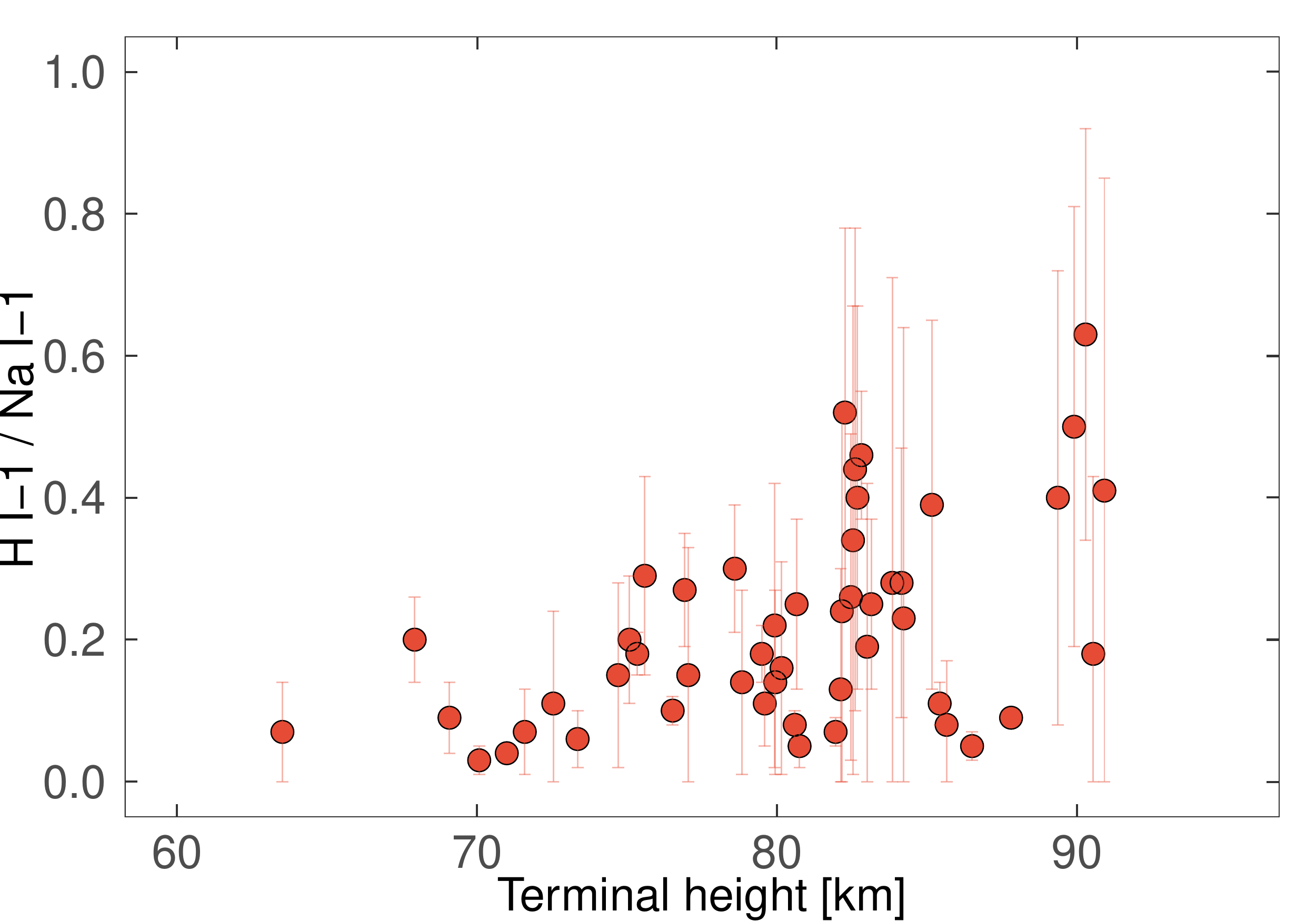}}
\caption{The observed H$\alpha$/Na I-1 intensity ratio as a function of meteor terminal height for meteoroids with detected H$\alpha$ emission.} 
\label{HaNa_vs_HE}
\end{figure}

Next, we studied the material strengths of all meteoroids with detected H emission, to see whether the suggested increased content of water and other H-bearing compounds (including organic matter) provides implications for the physical properties of these meteoroids. Our results in previous sections suggest that H emission is common in most cometary meteoroids, assuming that the meteor ablation temperatures are sufficient and that the meteoroid volatiles were not depleted by space weathering. Meteoroids with spectral signs of depleted volatile content were previously found to have increased material strengths, which appear to be related with material compaction during solar approaches \citep{2005Icar..174...15B, 2019A&A...629A..71M, 2021A&A...654A.147S}.  

The variations of estimated material types, based on the determined $P_{E}$ coefficients, for meteoroids with detected H emission are displayed in Fig. \ref{TJPE}. The dataset was found to include type IIIB fireballs (assumed to represent very fragile cometary materials), type IIIA fireballs (standard cometary materials), as well as numerous type II fireballs, which are assumed to have strength similar to carbonaceous materials. No Type I fireballs (representing the strongest material, characteristic for asteroidal debris) with H emission were found. This result is rather expected, as the vast majority of observed meteoroids with H emission originated from Halley-type cometary orbits.

When we look at the variations of the measured H/Na intensity ratios as a function of the meteoroid terminal heights (Fig. \ref{HaNa_vs_HE}), we recognize a trend suggesting that meteoroids with higher H content completely disintegrate higher up in the atmosphere. This would confirm the expected effect of increased volatile content on the resulting lower bulk strength of the meteoroid.

\subsection{Hydrogen emission in laboratory ablated meteorites} \label{sec:laboratory}

Our survey of meteor spectra have shown that variations of hydrogen emission can be detected mainly in cometary meteoroids. We assume that the detection of hydrogen in slower, mainly asteroidal meteoroids ($v_i <$ 30 km s\textsuperscript{-1}), is limited by the typically lower ablation temperatures which result in a weak H$\alpha$ line, often lost in the background noise. It can, however, also be predicted that on average, lower H content will be found within asteroidal bodies in comparison to cometary meteoroids, which are rich in ices, hydrated minerals and organic compounds. To better resolve the H emission from asteroidal materials, we studied high-resolution Echelle spectra of different meteorites obtained during their simulated ablation in plasma wind tunnel facilities. 

All of the analyzed meteorites represent asteroidal material and include a range of classes of carbonaceous chondrites, ordinary chondrites and achondrites. Some of the tested meteorite samples were omitted from this analysis, as the radiating plasma was found to contain additional source of H, possibly from a water vapor leaked from the internal cooling system. Fortunately, the contamination can be easily recognized, as the H emission is visible in the free plasma flow before the meteorite ablation starts. In all of the meteorites presented here, the H emission was only observed directly after the meteorite started to ablate and was not present in the plasma flow itself. With two exceptions, all of the ablated meteorites were meteorite falls. Meteorite falls, when available, were intentionally selected for this project, so that their composition is not significantly affected by terrestrial weathering. However, all sources of potential terrestrial contamination cannot be fully ruled out, including contamination during sample preparation (i.e., meteorite samples were cut with water and cleaned with ethanol), but considering that the spectra were obtained from a full ablation of a $\sim$ cm sized meteorite, their contribution to the analyzed spectra are expected to be minor for meteorite falls. 

\begin{figure}
\centerline{\includegraphics[width=\columnwidth,angle=0]{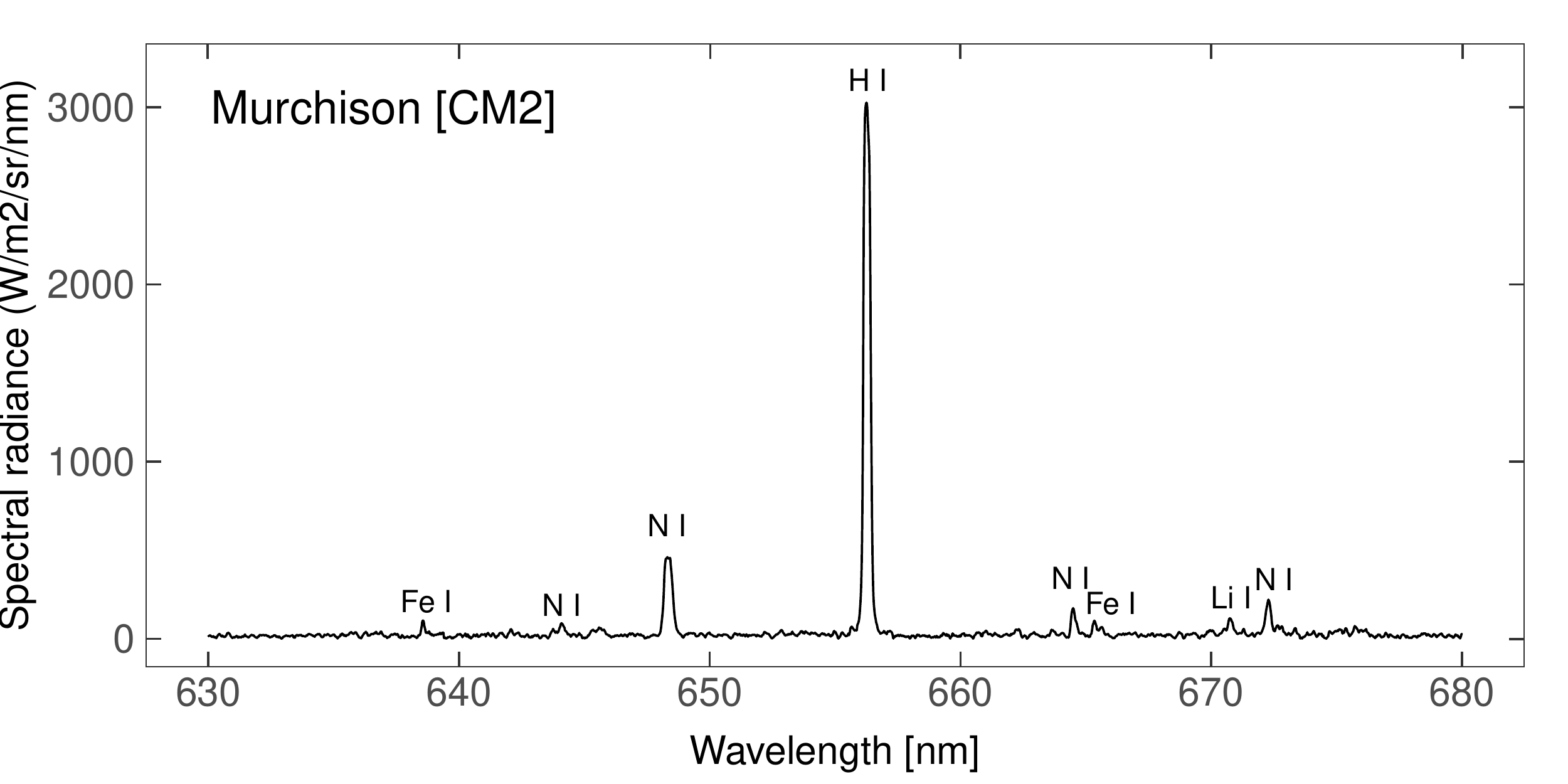}}
\centerline{\includegraphics[width=\columnwidth,angle=0]{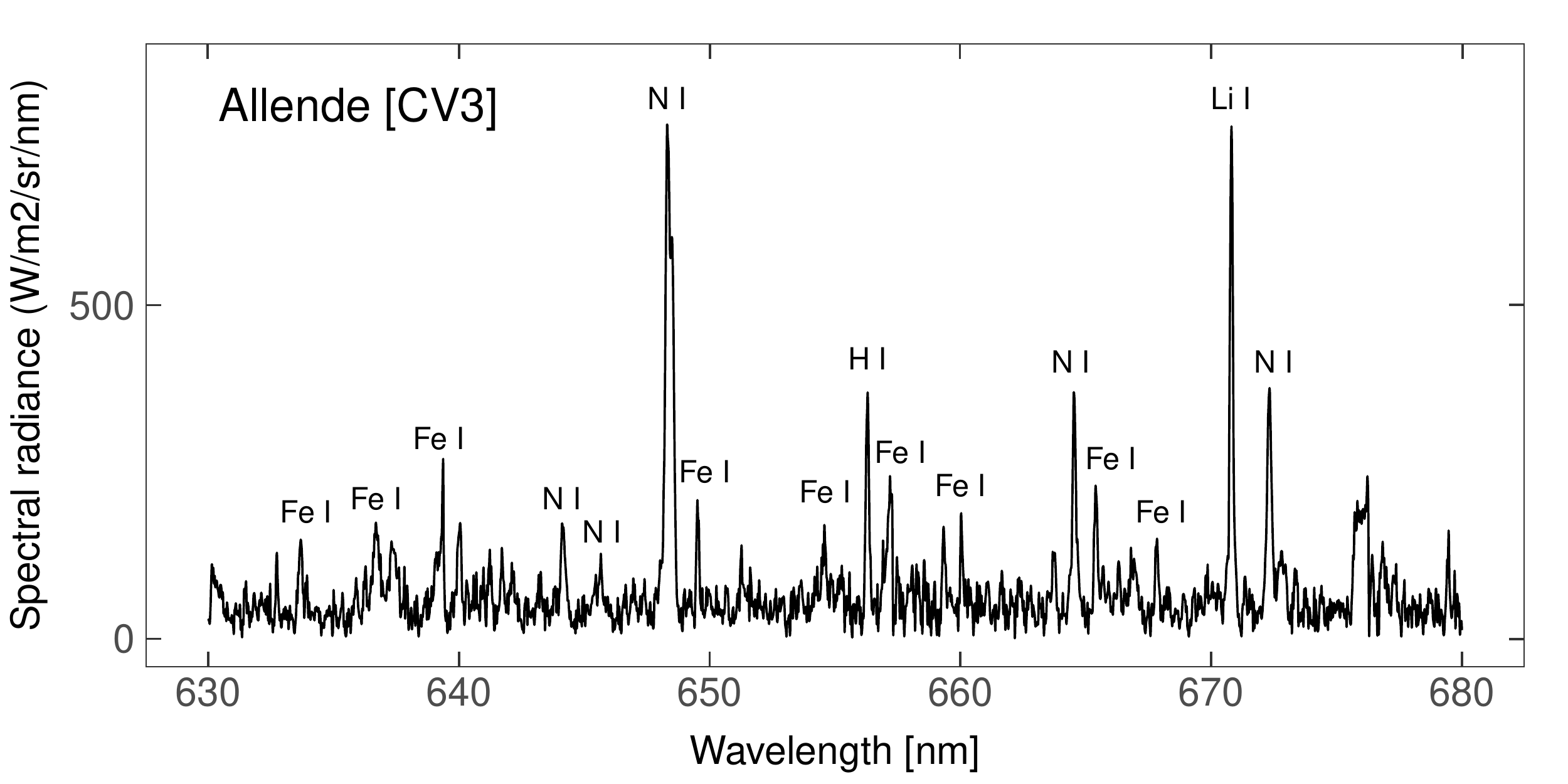}}
\centerline{\includegraphics[width=\columnwidth,angle=0]{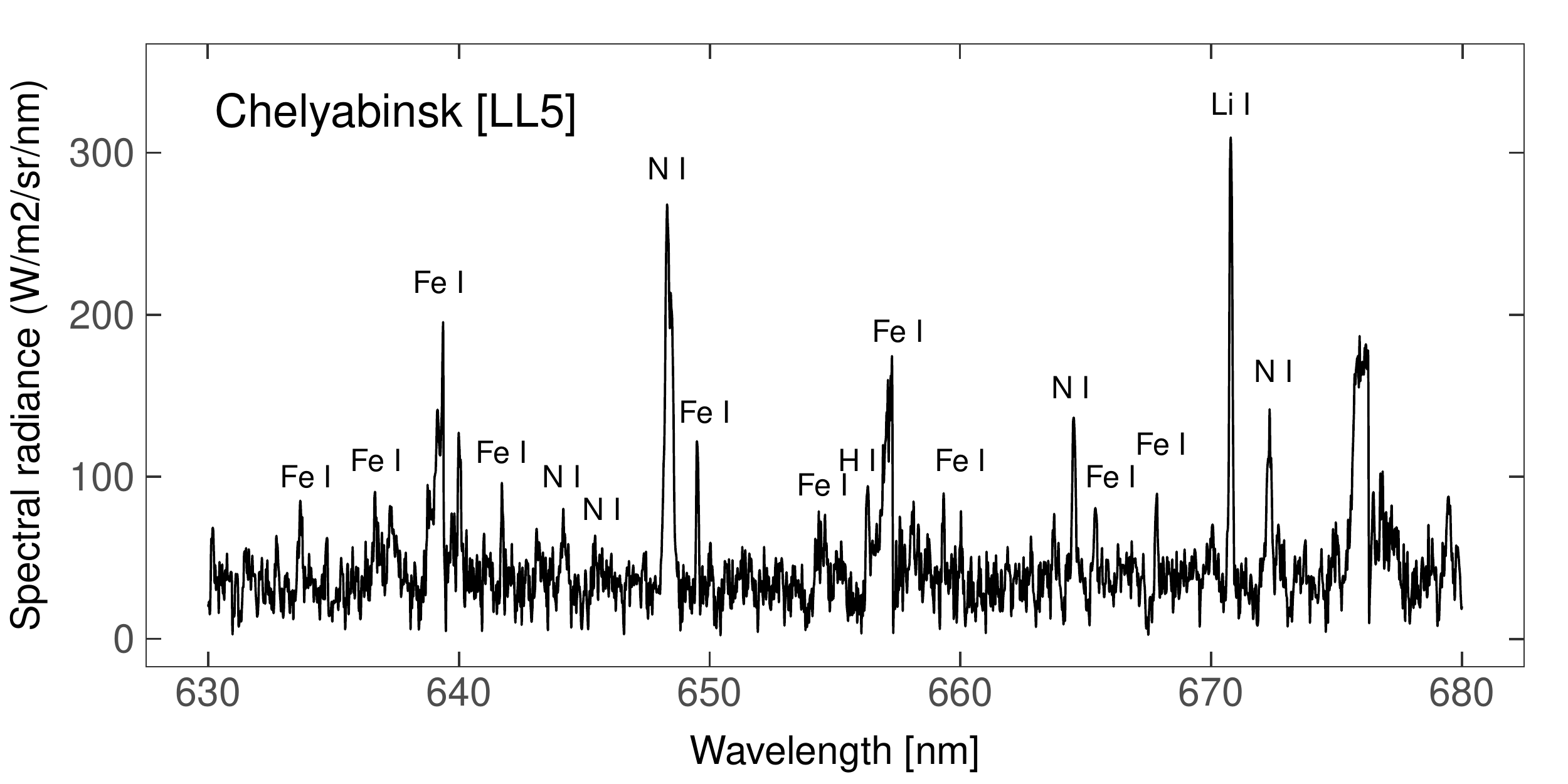}}
\caption{The calibrated emission spectra of three meteorites in the 630 - 680 nm region demonstrating the relative intensities of the H$\alpha$ line near 656.3 nm in meteorites of different types.} 
\label{MeteoriteProfiles}
\end{figure}

We also note that the studied meteorites already underwent ablation during their atmospheric entry, which may have caused some compositional alteration, particularly with respect to their volatile content. \citet{2004MNRAS.348..802T} have found Na overabundance in meteoroids compared to CI chondrites, and linked it to their loss during atmospheric heating. For our purposes here, we are mainly interested in the detection and relative intensity of the H emission in different meteorite types, and do not analyze in detail the differences in relative elemental abundances between meteoroids and meteorites, which will be the subject of our future studies. Examples of the emission spectra in the 630 - 680 nm region for three meteorites with various H$\alpha$ intensity are displayed in Fig. \ref{MeteoriteProfiles}. The Echelle meteorite spectra have $\sim$ 10-times higher resolution than the observational AMOS-Spec-HR data, which allowed accurate fit of the spectra, resulting in low uncertainty of the determined intensity ratios (Table \ref{Meteorite_values}).

\begin{figure}
\centerline{\includegraphics[width=\columnwidth,angle=0]{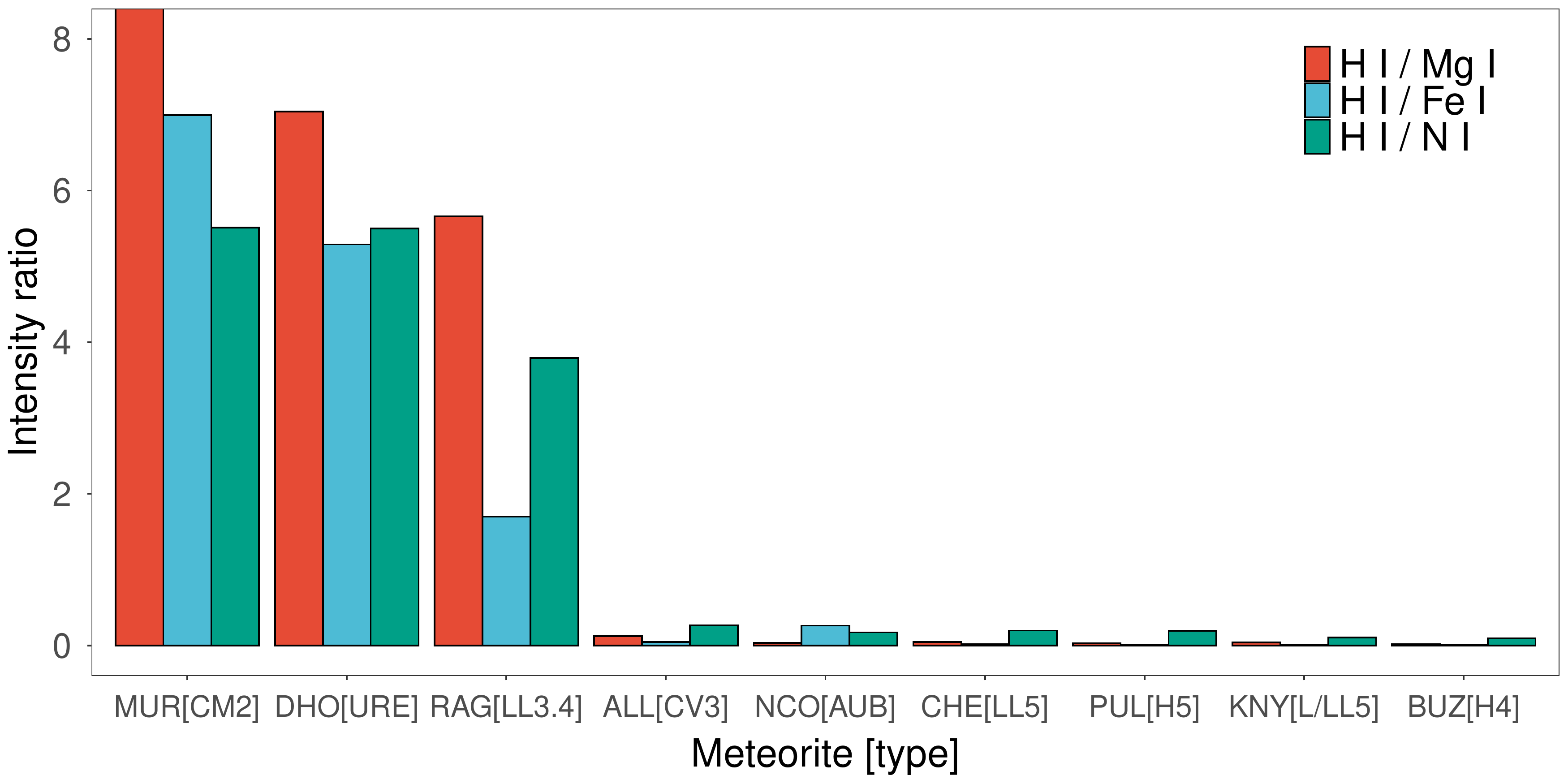}}
\caption{Measured relative H$\alpha$ intensity ratios in the emission spectra of meteorites ablated in a plasma wind tunnel under conditions representative of a low-speed meteor. The presented meteorites are in the order of the strongest H emission: MUR - Murchison (CM2), DHO - Dhofar 1575 (Ureilite), RAG - Ragland (LL3.4), ALL - Allende (CV3), NCO - Norton County (Aubrite), PUL - Pultusk (H5), CHE - Chelyabinsk (LL5), KNY - Knyahinya (L/LL5), BUZ - Buzzard Coulee (H4). The H I / Mg I intensity ratio for Murchinson is outside the plot, near 21. The displayed intensity ratios were derived by measuring emission multiplet ratios of HI-1/MgI-2, HI-1/FeI-15, HI-1/NI-21. These emission multiplets were selected as they are among the most prominent and well measured even in lower-resolution meteor spectra.} 
\label{Meteorites}
\end{figure}

The H$\alpha$ intensity ratios in the ablated meteorite samples, measured relative to the emission multiplets frequently studied in meteors (Section \ref{sec:survey}) are displayed in Fig. \ref{Meteorites} and Table \ref{Meteorite_values}. The H$\alpha$/Mg I-2 and H$\alpha$/Fe I-15 ratios are more affected by the specific meteorite composition. One can note the effects of the increased Fe content in the Ragland meteorite or high Mg content in Norton County (Fig. \ref{Meteorites}). The N lines in both meteor and meteorite spectra originate dominantly from the heated atmosphere and the generated plasma respectively \citep{1998SSRv...84..327C}, so their intensity is not strongly affected by the different meteorite composition. Overall, the strongest H emission was detected in the carbonaceous CM2-type meteorite Murchison, followed by the Ureilite achondrite Dhofar 1575, ordinary chondrite LL3.4 Ragland, carbonaceous CV3 meteorite Allende, and the aubrite achondrite Norton County (Table \ref{Meteorite_values}). The H emission was faint or close to the detection limit in the remaining meteorites (four ordinary chondrites of type H4, H5, L/LL5 and LL5).

\begin{figure}
\centerline{\includegraphics[width=\columnwidth,angle=0]{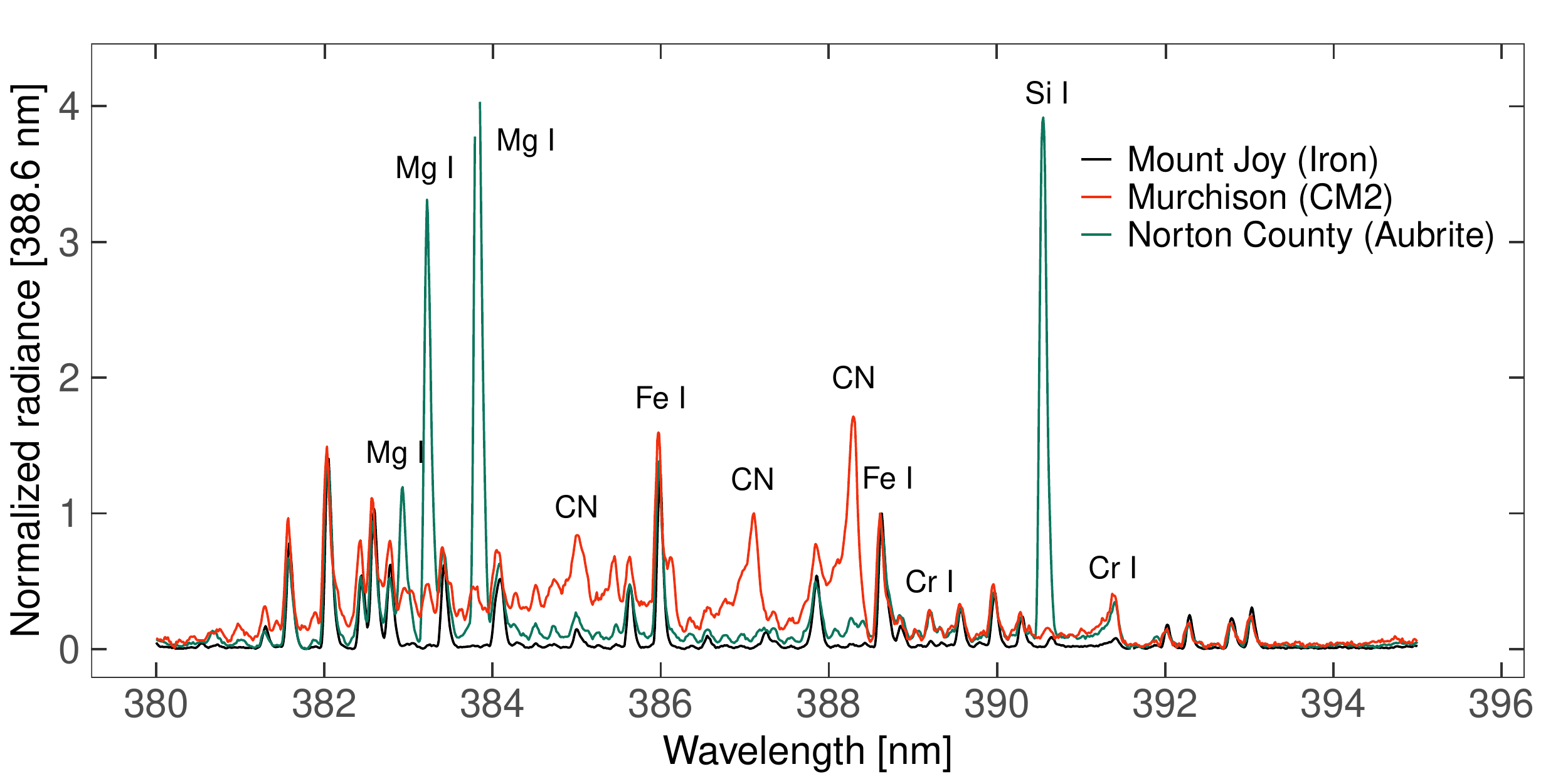}}
\caption{The calibrated emission spectra of three meteorites representing different compositional types in the 380 - 395 nm region. Different spectral signatures are demonstrated in this region depending on the meteorite contents of organic compounds, silicates and iron. The majority of unmarked lines belong to Fe I. The spectra are normalized to unity at the 388.6 nm Fe I line.} 
\label{CNcomp}
\end{figure}

These results well reflect the real bulk elemental composition of the tested meteorites (i.e., in the case of Dhofar 1575 and Ragland meteorite, both finds, the effect of terrestrial weathering should also be taken into account, see discussion below), and therefore confirm the potential of the H$\alpha$ line to trace the water and organic matter content of meteoroids. The high-resolution Echelle spectra of ablating meteorites also allowed us to analyze the CN band intensity, which directly points to the meteorite organic matter content. Previous attempts to detect CN in meteor spectra gave negative results \citep{1998EM&P...82...71R, 2004AsBio...4...67J, 2016Icar..278..248B}, due to strong contamination of the surrounding Fe lines and likely insufficient carbon content in studied meteoroids. Fig. \ref{CNcomp} shows CN emission in the Murchison meteorite and demonstrates different spectral signatures in the 380-395 nm region for meteorites with different compositional type. In this region, spectra of ordinary chondritic or iron-enhanced bodies are typically dominated by Fe I lines, while Mg I and Si I lines are dominant in the silicate-rich and Fe-poor Aubrite Norton County, and strong emission of CN can be seen in the organic-rich Murchison (CM2).

\begin{figure}
\centerline{\includegraphics[width=\columnwidth,angle=0]{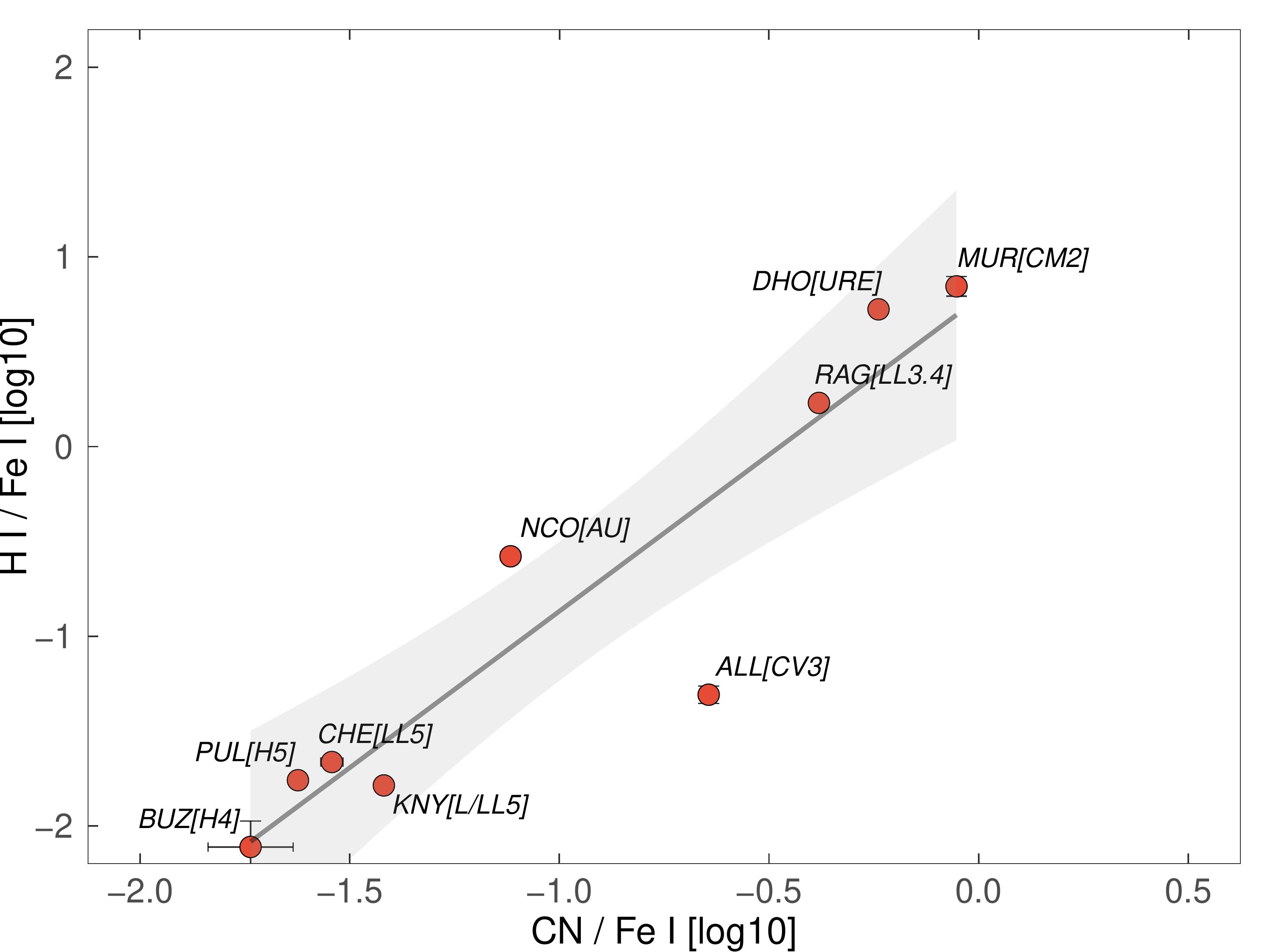}}
\centerline{\includegraphics[width=\columnwidth,angle=0]{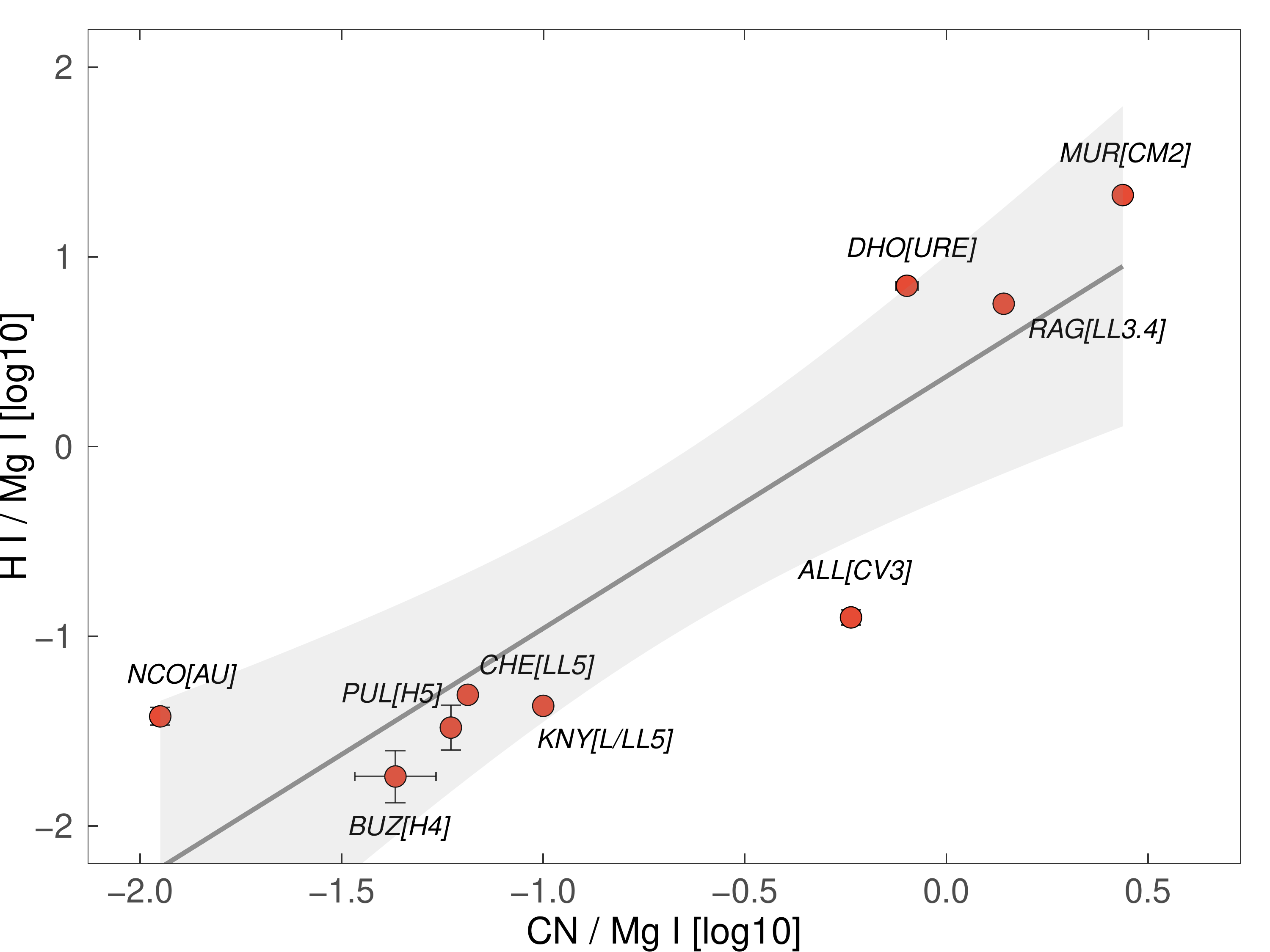}}
\caption{The dependency between the intensities of the H$\alpha$ line and CN band peak near 388.3 nm, measured relative to the MgI-2 and FeI-15 intensity, in the emisson spectra of meteorites ablated in a plasma wind tunnel under conditions representative of a low-speed meteor. The black line and gray area show the linear fit to the running average with standard error of the mean. The presented meteorites are the same as in Fig. \ref{Meteorites}. All of the tested meteorites are plotted for context, but the H$\alpha$ and CN intensities in the Pultusk, Buzzard Coulee, Chelyabinsk and Knyahinya meteorites were too low for their reliable confirmation and may instead represent faint background Fe lines. The significant difference in the relative H$\alpha$ intensity in the Norton County meteorite is caused by its Mg-rich and Fe-poor composition.} 
\label{HCN}
\end{figure}

We have found clear correlation between the intensities of H$\alpha$ and CN in the meteorite spectra (Fig. \ref{HCN}), confirming that the H$\alpha$ line corresponds with the increased volatile content in meteoroids. We note that while all of the tested meteorites are displayed in Fig. \ref{HCN} for better context, both H$\alpha$ and CN, considering their very low intensities (log\textsubscript{10}(CN/FeI) $<$ -1.5)), were not reliably detected in the Buzzard Coulee (H4), Pultusk (H5), Chelyabinsk (LL5) and Knyahinya (L/LL5) meteorites. 

The strongest emission of both H and CN was found in the CM2 carbonaceous chondrite Murchison. Murchison is a well studied meteorite, known for its high content of organic compounds, including amino-acids and hydrocarbons, as well as $\sim$ 10 wt.\% water content \citep{1970Natur.228..923K, 1997Natur.389..265E, 2018GeCoA.239...17B}. Due to the high intensity of the H$\alpha$ line in Murchison, the line was saturated in most frames of the Echelle spectrum, which resulted in the increased uncertainty of the calculated relative intensity ratios (Table \ref{Meteorite_values}). By comparison, the other carbonaceous chondrite studied in this work -- Allende (CV3) contains on average $<$ 1 wt.\% water content \citep{2018GeCoA.239...17B}, which was manifested in the relatively faint, though still recognizable, H$\alpha$ line in its emission spectrum (Fig. \ref{MeteoriteProfiles}).

The second strongest H emission was detected from the ureilite Dhofar 1575. Ureilites are rich in carbon in the form of graphite and nanodiamonds, and appear to have water content comparable with angrites and eucrites \citep{2015M&PS...50..782G, 2018LPI....49.2223D}. Besides Ragland, Dhofar 1575 was the only meteorite find (not fall) in our sample suite. While the weathering grade of the meteorite is low \citep{2014M&PS...49E...1R}, partial influence of terrestrial weathering on the detected H content cannot be fully disregarded.

Strong H emission was also observed from the LL3.4 ordinary chondrite Ragland, and also corresponded with high CN band intensity (Fig. \ref{HCN}). Ragland is however, out of the samples analyzed in this work, the meteorite most affected by terrestrial weathering, which has altered metallic Fe, Ni and troilite to iron oxides and hydroxides \citep{1986Metic..21..217R}. Considering that H and CN emission were not detected in any of the remaining four ordinary chondrites (Chelyabinsk, Knyahinya, Buzzard Coulee and Pultusk), one could assume that the moderate weathering of Ragland may have affected its composition and consequently the observed spectral features. Nevertheless, the mineralogical and chemical composition analyses of Ragland revealed some unusual features, such as the heavy oxygen isotopic composition and relatively high water content for ordinary chondrites \citep{1986Metic..21..217R}. It is therefore also plausible that the observed spectrum reflects Ragland's original, unusual properties. We note that the Ragland meteorite is also the least metamorphosed ordinary chondrite that was investigated.

Finally, the H$\alpha$ emission was also clearly confirmed, although fainter, in the aubrite meteorite Norton County. The determined H/Mg intensity ratio is low due to the high Mg I intensity in the spectrum of this meteorite, while the H/Fe intensity ratio is relatively high (Fig.\ref{HCN}). The varying relative H intensity in Norton County reflects its Mg-silicate-rich and Fe-poor composition \citep{2011M&PS...46..284H, 1985Metic..20..571E}.

\begin{table}
\centering
\small\begin{center}
\caption {Measured relative intensity ratios of H$\alpha$ in high-resolution spectra of ablated meteorites. Multiplet numbers are based on \citet{1945CoPri..20....1M}.} 
\resizebox{\columnwidth}{!}{
\begin{tabular}{lcrrr}
\hline\hline
\multicolumn{1}{l}{Meteorite}& %
\multicolumn{1}{c}{Class}& %
\multicolumn{1}{c}{HI-1/MgI-2} & %
\multicolumn{1}{c}{HI-1/FeI-15} & %
\multicolumn{1}{c}{HI-1/NI-21}  %
\\
\hline
Murchison      & CM2     & 21.18 $\pm$ 0.98  & 6.99 $\pm$ 0.84 & 5.51 $\pm$ 0.52 \\
Dhofar 1575    & Ureilite & 7.04 $\pm$ 0.05  & 5.29 $\pm$ 0.11 & 5.50 $\pm$ 0.15 \\
Ragland        & LL3.4    & 5.66 $\pm$ 0.20  & 1.70 $\pm$ 0.06 & 3.79 $\pm$ 0.23 \\
Allende        & CV3      & 0.13 $\pm$ 0.01  & 0.05 $\pm$ 0.01 & 0.27 $\pm$ 0.02 \\
Norton County  & Aubrite  & 0.04 $\pm$ 0.01  & 0.26 $\pm$ 0.01 & 0.18 $\pm$ 0.03 \\
Chelyabinsk    & LL5      & 0.05 $\pm$ 0.01  & 0.02 $\pm$ 0.01 & 0.20 $\pm$ 0.02 \\
Pultusk        & H5       & 0.03 $\pm$ 0.01  & 0.01 $\pm$ 0.01 & 0.20 $\pm$ 0.08 \\
Knyahinya      & L/LL5    & 0.04 $\pm$ 0.01  & 0.02 $\pm$ 0.01 & 0.11 $\pm$ 0.01 \\
Buzzard Coulee & H4       & 0.02 $\pm$ 0.01  & 0.01 $\pm$ 0.01 & 0.10 $\pm$ 0.03 \\
\hline
\end{tabular}}
\label{Meteorite_values}
\end{center}
\end{table}

\section{Discussion} \label{sec:discussion}

Overall, we report the detection of H emission in five out of nine analyzed meteorites. These include two carbonaceous chondrites (strong emission from the CM2 Murchison, fainter emission from the CV3 Allende), a ureilite, an unusual, weathered LL3.4 ordinary chondrite Ragland, and an aubrite achondrite Norton County. Our results suggest that H emission can be observed from asteroidal meteoroids, and appears to be preferred from carbonaceous chondrites and achondrites with enhanced organic matter content. 

One issue that remains to be addressed is the process of excitation of the atomic H in the ablated meteorites. The performed laboratory simulated ablation recreates conditions representative of a slower meteor (assumed 10 to 12 km s\textsuperscript{-1}) at an altitude of $\sim$ 80 km \citep{2017ApJ...837..112L}. The results from our meteor survey (Fig. \ref{absMag_vi_Ha_nonHa}) however suggest that no H emission is detected in meteors slower than $\sim$ 30 km s\textsuperscript{-1}. We also note that the spectra of ablated meteorites do not contain other lines typically associated with the high temperature component in meteors \citep{1994P&SS...42..145B} (with the exception of N I lines, discussed in the next paragraph). High-temperature lines of Si II near 634.7 and 637.1 nm, commonly observed in faster meteors, were not detected in the ablating meteorite spectra. In the meteor survey presented here, all spectra with the visible H$\alpha$ line also had Si II emission present. The Si II lines have lower excitation energy (10.07 eV) than H$\alpha$, and considering the overall high content of Si in most meteorites, should be visible. Similar discrepancy was observed in a Geminid spectrum by \citet{2004AsBio...4..123J}, where as a possible explanation, it was argued that non-evaporation of the whole amount of the silicon, or a smaller fraction of silicon in an ionized state not corresponding to high-temperature spectrum component, could occur. 

To estimate the temperatures of the radiating plasma, we have fitted high-resolution meteorite spectra with a radiative transfer model. The model considers a local thermal equilibrium and optically thick radiating plasma with self-absorption, similar to the successfully tested model of \citet{1993A&A...279..627B}. We have found that the observed spectra dominated by the Fe lines fit well with temperatures between 4600 - 6000 K, depending on the individual experiment/meteorite sample. At these temperatures, the relative presence H$\alpha$ line should be quite small, below the observed values. We suggest that the plasma flow generated by the setup does not follow the thermal equilibrium for all components and that an alternative process of the H excitation occurred during the simulated ablation. The only other detected high-temperature lines in meteorite spectra were of N I (Fig. \ref{MeteoriteProfiles}). Given that the present N atoms dominantly originate from the generated incoming high-enthalpy plasma, the present N I lines may have also been excited by a non-thermal process.

Even though a different dominant excitation process of H may occur in meteors and ablated meteorites in the plasma wind tunnel, our results point out relevant differences in spectral properties of different types of asteroidal material and can help to interpret observed meteor spectra with no reference to the real meteoroid composition. Given that ordinary chondrites did not exhibit any H emission, while its intensity appears to be significantly increased in some carbonaceous chondrites and achondrites, the potential detection of H emission in spectra of asteroidal meteoroids could be used to constrain the unknown composition of observed meteoroids. The collection and quantification of more emission spectra of meteorites of different compositional types would help to better constrain such diagnostic methods. 

\section{Conclusions} \label{sec:conclusions}

We present the first large-scale survey of hydrogen emission from meteors and ablated meteorites. Our results suggest that the meteor H emission correlates with the meteoroid volatile (Na and CN) content, and likely presents a suitable tracer of H\textsubscript{2}O molecules (as ice and bound in minerals) and organic compounds in meteoroids. Due to the high excitation of the H$\alpha$ line, it is favored to be detected in meteors which form the high-temperature spectral component (mainly $v_i >$ 50 km s\textsuperscript{-1}). H emission was not detected in any meteoroids slower than 30 km s\textsuperscript{-1}. In the context of meteor observations, the H$\alpha$ line is clearly a characteristic spectral feature of cometary meteoroids.

We have found that $\sim$ 92\% of all meteoroids with detected H emission originated from Halley-type and long-period cometary orbits (T\textsubscript{J} $<$ 2). This sample was represented by meteoroids from several major meteoroid streams including Perseids, Leonids, Lyrids, Orionids, sigma Hydrids and few other minor streams. The cometary meteoroids with the highest relative H intensities originated from orbits with T\textsubscript{J} $<$ 1. Only three meteors with detected H emission were found to originate from Jupiter-family comet type orbits and asteroidal orbits (T\textsubscript{J} $>$ 2) and included a Geminid and a northern Taurid. These results give validity to the hypotheses that some regions of the surfaces of Phaethon and comet 2P/Encke may still be hydrated. Our results also suggest that hydrogen is being depleted from meteoroids with close perihelion approaches (q $<$ 0.4 au) to the Sun.

Hydrogen was not detected in any asteroidal meteoroids in our survey. Using emission spectra of laboratory ablated meteorite samples, we have however found that H emission can occur from asteroidal material, and apparently correlates with the water and organic matter content of these meteorites. Out of the tested samples, the most notable H emission was detected from carbonaceous chondrites (CM2 and CV3) and achondrites (a ureilite and an aubrite). Spectra of ordinary chondrites did not exhibit H emission, with the only exception of the moderately weathered LL3.4 chondrite Ragland. The process of excitation of H in spectra of meteorites ablated in the plasma wind tunnel was likely different to the dominant process of thermal excitation in meteors, which typically produces a weak H$\alpha$ line. The detection of H emission in asteroidal meteors could be potentially used to identify meteoroids of carbonaceous or achondritic composition. 

\section*{Data availability}
The spectral and trajectory data of presented meteors and meteorites will be made available upon a reasonable request to the corresponding author.

\section*{Acknowledgements}

The authors are grateful to the High Enthalpy Flow Diagnostics Group (HEFDiG) team of the Institute of Space Systems, University of Stuttgart for the preparation and participation in the meteorite ablation experiments. The AMOS team acknowledges the support of staff contributing to the operation of the AMOS systems in Slovakia, and the Instituto de Astrofísica de Canarias and Institute for Astronomy, University of Hawaii for providing support with the installation and maintenance of AMOS systems in Canary Islands and Hawaii. PM and JT acknowledge the help of Dr. Pavel Vojtek with the calibration of AMOS-Spec cameras. This work was supported by the ESA contract No. 4000128930/19/NL/SC, the Slovak Research and Development Agency grant APVV-16-0148, the Slovak Grant Agency for Science grant VEGA 1/0218/22 and the Comenius University Grant G-21-193-00. The authors are grateful to Jiří Borovička for his review and comments that helped improve this manuscript.
     
\bibliographystyle{mnras}
\bibliography{references}
\bsp	
\label{lastpage}
\end{document}